\documentclass[a4paper,12pt]{article}
\usepackage{amsmath}
\usepackage{amsfonts}
\usepackage{amssymb}
\usepackage{epsfig}
\usepackage{psfrag}

\def\fmfframe(#1,#2)(#3,#4)#5{%
  \leavevmode
  \hbox{\vbox{\vskip#2\unitlength\par
              \hbox{\hskip#1\unitlength#5\hskip#3\unitlength}\par
              \vskip#4\unitlength}}}

\newcommand{\clap}[1]{\makebox[0pt][c]{#1}}
\newcommand{\vect}[1]{{\boldsymbol{#1}}}

\newcommand{\itext}[1]{\quad\text{#1}\quad}

\newcommand{\union}{\mathop{\textstyle\bigcup}\limits}

\newcommand{\defd}{\equiv}
\newcommand{\PO}{\mathrm{P}}
\newcommand{\on}{\mathclose{\mathchoice{\bigr|}{|}{|}{|}}}
\newcommand{\clbrack}{\mathopen{\mathchoice{\bigl[}{[}{[}{[}}}
\newcommand{\crbrack}{\mathclose{\mathchoice{\bigl]}{]}{]}{]}}}
\newcommand{\windon}[3]{#1 \clbrack #3\on_{#2} \crbrack}
\newcommand{\alphaTD}{\windon{\alpha_{D}}{T}{\hphi}}

\newcommand{\deghphiD}{\windon{\deg}{D}{\hphi}}

\newcommand{\degDbeta}{\windon{\deg}{D^\beta}{\hphi}}

\newcommand{\half}{\tfrac{1}{2}}

\newcommand{\eps}{\varepsilon}
\newcommand{\thet}{\vartheta}
\newcommand{\dagg}{{\mathord{+}}}
\newcommand{\ensp}{\enspace}

\newcommand{\Sph}{S}
\newcommand{\B}{B}
\newcommand{\SU}{{\mathrm{SU}}}

\newcommand{\U}{{\mathrm{U}}}
\newcommand{\su}{{\mathfrak{su}}}
\newcommand{\R}{{\mathbb{R}}}
\newcommand{\Z}{{\mathbb{Z}}}
\newcommand{\dd}{{\mathrm{d}}}
\newcommand{\ee}{{\mathrm{e}}}
\newcommand{\ii}{{\mathrm{i}}}
\newcommand{\tr}{\operatorname{tr}}

\newcommand{\openone}{1}

\newcommand{\calC}{{\mathcal{C}}}
\newcommand{\calM}{{\mathcal{M}}}

\newcommand{\south}{s}

\newcommand{\hphi}{{\hat{\phi}}}

\newcommand{\hx}{{\vect{\hat x}}}
\newcommand{\ha}{{\vect{\hat \alpha}}}

\newcommand{\Udagg}{U^\dagg}

\newcommand{\nab}[1]{n_{#1=#1_0}}
\newcommand{\nabt}[1]{n_{#1=#1_0}^{\thet=\pi}}

\begin{document}

\begin{flushright}
FAU-TP3-99/6\\
hep-th/9909004
\end{flushright}

\vspace{5mm}

\begin{center}
{\Large Instantons and Monopoles in General Abelian Gauges}

\vspace{1cm}

{O.~Jahn%
\footnote{E-mail: \texttt{jahn@theorie3.physik.uni-erlangen.de}}}

\textit{Institut f\"ur Theoretische Physik\\
Friedrich-Alexander-Universit\"at Erlangen-N\"urnberg\\
Staudtstra{\ss}e 7, D-91058 Erlangen, Germany}
\end{center}

\vspace{2.5mm}

\begin{abstract}
  A relation between the total instanton number and the quantum-numbers
  of magnetic monopoles that arise in general Abelian gauges in
  $\SU(2)$ Yang-Mills theory is established.  The instanton number is
  expressed as the sum of the `twists' of all monopoles, where the
  twist is related to a generalized Hopf invariant.  The origin of a
  stronger relation between instantons and monopoles in the Polyakov
  gauge is discussed.
\end{abstract}

\section{Introduction}

Instantons and (Abelian projection) monopoles are both topological
objects that are associated with low-energy phenomena in QCD.  While
instantons provide a solution to the $\U(1)$ problem
\cite{tHooft:1976a,tHooft:1976b} and an explanation for chiral symmetry
breaking \cite{Shuryak:1988}, they have not been able to explain color
confinement yet \cite{Callan:1978gz,Chu:1994vi}.  A possible mechanism
for the latter is the dual Meissner effect due to condensation of
magnetic monopoles that arise in so-called Abelian gauges
\cite{Mandelstam:1974pi,'tHooft:1981ht}.  Lattice simulations indicate
that magnetic monopoles indeed play an important role for confinement
\cite{Kronfeld:1987ri,Suzuki:1990gp,Stack:1994wm,Bali:1996dm}.  Since
lattice simulations also indicate that the transition to a deconfined
phase and the restoration of chiral symmetry occur at approximately the
same temperature, it would be puzzling if they were generated by
completely independent mechanisms.  There is indeed evidence from a
number of studies both in the continuum and on the lattice that
instantons and monopoles are correlated in several Abelian gauges (see,
e.g.,
\cite{Chernodub:1995tt,Brower:1997js,Brower:1998ep,Hart:1996wk,Bornyakov:1996wp,Sasaki:1998ww}).
A connection between the instanton number (Pontryagin index) and
magnetic charges has already been considered in Refs.\ 
\cite{Christ:1980,Gross:1981}.  The detailed relation between the total
instanton number and the quantum numbers of magnetic monopoles has so
far only been established in the Polyakov gauge (or the related
modified axial gauge) \cite{Reinhardt:1997rm,Ford:1998bt,Jahn:1998nw}.


In the standard model, Taubes has shown how monopole fields can be used
to generate topological charge \cite{Taubes:1984}.  As pointed out by
van Baal in Ref.\ \cite{vanBaal:1997}, similar arguments may be made in
the context of Abelian projection in pure Yang-Mills theory.  This has
been demonstrated explicitly for a new finite temperature instanton
(caloron) solution by Kraan and van Baal in Ref.\ \cite{Kraan:1998pm}.
There, it has been shown that the instanton number is carried by a
magnetic monopole that makes a full rotation in color space along its
(closed) world-line.  It has been noted that the relevant topology for
this `twist' of the monopole is the Hopf fibration.  These observations
are worked out in greater detail for general configurations in the
present work.

Section \ref{sec:genab} presents a short review of the definition of
general Abelian gauges in terms of an auxiliary Higgs field and of the
characterization of the magnetic monopole singularities arising in
these gauges.  In Sec.\ \ref{sec:inst-genab}, the general relation
between the instanton number and the auxiliary Higgs field is
established for the Euclidean `space-time' $\Sph^4$.  Section
\ref{sec:gen-hopf} provides a generalization of the Hopf invariant of
maps from $\Sph^3$ to $\Sph^2$ to maps from $\Sph^2\times\Sph^1$ to
$\Sph^2$.  This invariant is used in Sec.\ \ref{sec:loop} to derive the
contribution of a single monopole loop to the instanton number.  The
resulting relation between the instanton number and the generalized
Hopf invariants of monopoles is illustrated with the example of a
single-instanton solution that is known to lead to a monopole loop in
the (differential) maximal Abelian gauge \cite{Brower:1997js}.  In
Sec.\ \ref{sec:nontriv}, the contribution of topologically non-trivial
monopole loops to the instanton number on the space-time
$\Sph^3\times\Sph^1$ is derived.  Section \ref{sec:polyakov-gauge}
gives a qualitative explanation for the existence of a stronger
relation between instantons and monopoles in the Polyakov gauge.  The
final section contains a discussion of the results.

\section{Monopoles in general Abelian gauges}
\label{sec:genab}

Throughout this work, we consider pure $\SU(2)$ Yang-Mills theory.  The
term `Abelian gauge' will be used for gauges that are defined by the
diagonalization of some field $\phi[A]$ that transforms according to
the adjoint representation of the gauge group,
\begin{equation}
  \label{eq:ad-trafo}
  \phi(x) \to \Omega(x) \phi(x) \Omega^\dagg(x)
\end{equation}
under a gauge transformation $\Omega(x)\in\SU(2)$.  Due to this
property, we will call $\phi$ an \emph{auxiliary Higgs field}.  It is
not a fundamental field of the theory but rather a functional of the
gauge potential $A$.  The field $\phi$ can take values in either the
gauge group (in our case $\SU(2)$) or its algebra ($\su(2)$). 
Well-known examples are the Polyakov gauge, where $\phi$ is the
(time-dependent) Polyakov line,
\begin{equation}
  \label{eq:pline}
  \phi(\vect x,t) = \PO\exp \biggl(
  \int_t^{t+\beta}\!\!\dd t'\, A_0(\vect x,t')
  \biggr) \in \SU(2) 
\end{equation}
and the maximal Abelian gauge where
$\phi=\vect\phi\cdot\vect\sigma\in\su(2)$ minimizes the functional
\begin{equation}
  \label{eq:RmAg}
  R[\phi,A] = \int \!\dd^4 x \, \tr \left(
    [ \partial_\mu + A_\mu , \phi ]^2
  \right)
\end{equation}
under the constraint $|\vect\phi|=1$.

Monopole singularities arise where $\phi$ does not define a direction
in color space, i.e., where $\phi=0$ for $\phi\in\su(2)$ or
$\phi=\pm\openone$ for $\phi\in\SU(2)$.  Since these conditions involve
three equations, the monopole singularities will generically occupy
points in three-dimensional space or one-dimensional submanifolds
(world-lines) in four-dimensional space-time.  Around these points, the
direction of the auxiliary Higgs field defines a map from a
two-dimensional sphere $\Sph^2$ to another $\Sph^2$.  (In space-time,
one has to consider two-spheres that link with the monopole
world-line.)  The winding number of this map provides the charge of the
magnetic monopole singularity that appears in the diagonal part of the
gauge potential after gauge fixing.  It can be expressed as
\begin{equation}
  \label{eq:m-deg-phi}
  m = \deg[\vect\hphi] =
  \frac{1}{8\pi} \int_{\Sph^2} \epsilon_{i j k} \hphi_i \, \dd\hphi_j \wedge
  \dd\hphi_k
  = \frac{-\ii}{16\pi} \int_{\Sph^2} \tr \hphi\, \dd\hphi \wedge \dd\hphi \ensp,
\end{equation}
where the unit vector $\vect\hphi$ in the direction of the Higgs field
is defined via the relations
\begin{equation}
  \label{eq:hphi}
  \begin{aligned}
    \phi &= \vect\phi\cdot\vect\sigma
    \itext{and}
    \vect\hphi = \frac{\vect\phi}{|\vect\phi|}
    &&\itext{for} \phi\in\su(2) \ensp, \\
    \phi &= \cos\beta + \ii\vect\hphi\cdot\vect\sigma \sin\beta
    &&\itext{for} \phi\in\SU(2) \ensp,
  \end{aligned}
\end{equation}
and $\hphi=\vect\hphi\cdot\vect\sigma$ denotes the corresponding
$\su(2)$ matrix.

Using the fact that the gauge fixing transformation $\Omega$
diagonalizes $\hphi$,
\begin{equation}
  \label{eq:phi-diag}
  \Omega \hphi \Omega^\dagg = \sigma_3 \ensp,
\end{equation}
$m$ can be expressed in terms of $\Omega$,
\begin{equation}
  \label{eq:m-Omega}
  m = \frac{\ii}{4\pi} \int_{\Sph^2} \tr \sigma_3
  (\dd\Omega\,\Omega^\dagg)^2 \ensp.
\end{equation}
Since the integrand is a total differential,
$(\dd\Omega\,\Omega^\dagg)^2 = \dd(\dd\Omega\,\Omega^\dagg)$, $\Omega$
has to be discontinuous at some point $x_1$ on $\Sph^2$ if $m\ne0$.
This is the origin of the Dirac string singularity in the Abelian
projected gauge potential.  Since the Higgs field is continuous on
$\Sph^2$, the discontinuity in $\Omega$ has to be Abelian,
\begin{equation}
  \label{eq:Omega-string}
  \Omega(x) \to \ee^{-\ii\psi(x)\sigma_3} \Omega_0   
  \itext{for}
  x \to x_1 \ensp.
\end{equation}
The magnetic charge can be expressed as the winding number of the phase
$\psi$ along an infinitesimal closed curve $\calC$ around $x_1$ on
$\Sph^2$,
\begin{equation}
  \label{eq:m-psi}
  m = \frac{1}{2\pi} \int_{\calC} \dd\psi \ensp.
\end{equation}

Note that although the above discussion does not directly apply to the
maximal Abelian gauge since the constraint $|\vect\phi|=1$ does not
permit zeros of $\vect\phi$, discontinuities of $\vect\hphi$ cannot in
general be avoided also in this gauge and monopole singularities arise
after gauge fixing.  In this case, of course, also the auxiliary Higgs
field itself is discontinuous.

\section{Instantons in general Abelian gauges}
\label{sec:inst-genab}

The above discussion shows that all information about the positions and
charges of the monopoles is present in the auxiliary Higgs field that
defines the Abelian gauge in question.  One is prompted to ask whether
information about the number of instantons is also included.  Since the
latter relates to global properties of the gauge field it is useful to
consider a specific space-time geometry.  For simplicity, we choose
$\Sph^4$.  It can be covered by two charts.  We will use one large
chart that covers all of $\Sph^4$ with the exception of one point and
as a second chart a small neighborhood of that point.  The excluded
point can be chosen such that the direction of the Higgs field is
well-defined on the small chart.  In the overlap, the gauge fields on
the two charts are related by a gauge transformation with a transition
function $U\in\SU(2)$,
\begin{equation}
  \label{eq:transition-function}
  A^{(1)} = \Udagg ( A^{(2)} + \dd ) U \ensp.
\end{equation}
Since the Higgs field transforms according to the adjoint
representation of the gauge group (it belongs to an associated fiber
bundle), the Higgs fields on the two charts are related by the same
gauge transformation,
\begin{equation}
  \label{eq:transition-higgs}
  \phi^{(1)} = \Udagg \phi^{(2)} U \ensp.
\end{equation}

We use stereographic projection to parameterize the large chart by
$\R^4$.  Equation~(\ref{eq:transition-function}) then turns into the
statement that $A^{(1)}$ approaches a pure gauge at infinity,
\begin{equation}
  \label{eq:pure-gauge}
  A^{(1)}(x) \sim \Udagg(\hat x) \,\dd U(\hat x)
  \itext{for}
  |x| \to \infty \ensp.
\end{equation}
We drop the superscript $(1)$ in the following because we do not need
the second chart any more.  The winding number (or degree) of $U$ as a
map from $\Sph^3$ to $\SU(2)\cong\Sph^3$ is the total instanton number
$\nu$ of $A$,
\begin{equation}
  \label{eq:nu}
  \nu = \deg[U] \equiv \frac{1}{24\pi^2} 
  \int_{\Sph^3} \tr \left[ (\Udagg \dd U)^3 \right]
\end{equation}
The Higgs field approaches the corresponding
gauge transform of a constant (the value of the Higgs field on the
excluded point of $\Sph^4$),
\begin{equation}
  \label{eq:pure-gauge-higgs}
  \phi(x) \to \phi_\infty(\hat x) 
  \defd \Udagg(\hat x) \, \phi_0 \, U(\hat x)
  \itext{for}
  |x| \to \infty \ensp.
\end{equation}
Due to our choice of charts, the direction of $\phi_\infty$ is
well-defined.  It provides a map from $\Sph^3$ to $\Sph^2$.  Such maps
fall into different homotopy classes and can be characterized by the
so-called \emph{Hopf invariant} (see e.g.\ \cite{Dubrovin}).  It is
usually defined in an indirect way: Let $\omega_2$ denote the volume
form on $\Sph^2$ (strictly speaking, the pull-back of it),
\begin{equation}
  \label{eq:vol-S2}
  \omega_2 = \half\epsilon_{i j k} \hphi_i\,\dd\hphi_j\wedge\dd\hphi_k
  = -\tfrac{\ii}{4} \tr\hphi\,\dd\hphi\wedge\dd\hphi \ensp.
\end{equation}
Since the second homology group of $\Sph^3$ is trivial, $H_2(\Sph^3)=0$
(`$\Sph^3$ does not contain non-contractible two-spheres'), $\omega_2$
is closed and can be written as a total derivative, $\omega_2=\dd
\upsilon$, where $\upsilon$ is a one-form.  The Hopf invariant is
defined as
\begin{equation}
  \label{eq:Hopf}
  \alpha[\hphi_\infty] 
  \equiv \frac{1}{16\pi^2} \int_{\Sph^3} \upsilon \wedge \dd \upsilon 
\end{equation}
and is independent of the choice of $\upsilon$.  Geometrically, the
Hopf invariant is given by the linking number of the preimages of two
arbitrary points on $\Sph^2$.  The preimages are generically
one-dimensional curves and have an orientation induced from the
neighborhood of the two points.  The linking number is defined as the
number of times one has to cross the two preimages to disentangle them
with orientations taken properly into account.  It has the algebraic
representation
\begin{equation}
  \label{eq:linknum}
  l = \frac{1}{4\pi} \int \frac{\vect x-\vect x'}{|\vect x-\vect x'|^3}
  \cdot \left( \dd\vect l \times \dd\vect l' \right)
\end{equation}
where the line integrals are performed over the two preimages.  One can
show that $l$ is independent of the choice of the two points on
$\Sph^2$.

The representation (\ref{eq:pure-gauge-higgs}) can be used to express
$\omega_2$ in terms of $U$,
\begin{equation}
  \label{eq:V-phi-inf}
  \omega_2 = \ii \tr[ \hphi_0 (\dd U \Udagg)^2 ] \ensp,
\end{equation}
which can be easily integrated,
\begin{equation}
  \label{eq:v}
  \omega_2 = \dd \upsilon
  \itext{with}
  \upsilon = \ii \tr[ \hphi_0\, \dd U \Udagg] \ensp.
\end{equation}
Without loss of generality we may choose $\hphi_0=\sigma_3$ yielding
\begin{equation}
  \begin{split}
  \upsilon \wedge \dd \upsilon &= -\tr[ \sigma_3 \,\dd U \Udagg]
  \wedge \tr[ \sigma_3 \,\dd U \Udagg 
  \wedge \dd U \Udagg] \\
  &= - (\dd U \Udagg)_3 
  \wedge \frac{\ii}{2} \epsilon_{3 i j} (\dd U \Udagg)_i
  \wedge (\dd U \Udagg)_j \\
  &= - \frac{\ii}{6} \epsilon_{i j k} (\dd U \Udagg)_i
  \wedge (\dd U \Udagg)_j \wedge (\dd U \Udagg)_k \\
  &= - \frac{2}{3} \tr \left[ (\dd U \Udagg)^3 \right] \ensp,    
  \end{split}
\end{equation}
where the anti-commutativity of the wedge product has been exploited.
We find that the Hopf invariant is given by the negative of the degree of $U$,
\begin{equation}
  \label{eq:hopf-vs-degree}
  \alpha[\hphi_\infty] 
  = - \deg[U] 
  = - \nu \ensp.
\end{equation}
The instanton number is therefore identical to the negative of the Hopf
invariant of the auxiliary Higgs field at infinity.

How does the latter relate to monopoles?  The necessity of points where
$\hphi$ is undefined for non-vanishing instanton number follows
immediately: a non-trivial $\hphi_\infty\colon\Sph^3\to\Sph^2$ cannot
be deformed into a constant continuously and is therefore not
extendable to $\R^4$.  The question whether these points are monopoles
(i.e., have non-zero magnetic charge) and how their charges relate to
the instanton number requires a more detailed analysis.

Before this, we investigate how the instanton number decomposes into
contributions from the individual monopoles.  Consider the generic case
of an arbitrary number of closed monopole loops in $\Sph^4$.  Since
loops cannot link in four dimensions, it is possible to enclose the
individual loops in disjoint four-volumes $V_i$ that are topologically
trivial (have no holes).  The Hopf invariant has the nice property of
being additive in the sense that $\alpha[\hphi_\infty]$ can be written
as the sum of the Hopf invariants of $\hphi$ on the boundaries of the
volumes $V_i$,
\begin{equation}
  \label{eq:hopf-sum}
  - \nu = \alpha[\hphi_\infty] 
  = \sum_i \windon{\alpha}{\partial V_i}{\hphi}
  \ensp,
\end{equation}
since $\hphi$ is continuous outside of the $V_i$.  Furthermore, since
(the adjoint of) the gauge fixing transformation $\Omega$ that
diagonalizes $\phi$ is related to $\hphi$ in the same way as $U$ to
$\hphi_\infty$,
\begin{equation}
  \label{eq:hphi-diag}
  \hphi = \Omega^\dagg \sigma_3 \, \Omega \ensp,
\end{equation}
the individual contributions are identical to the respective degrees of
$\Omega$,
\begin{equation}
  \label{eq:hopf-inst-indiv}
  \windon{\alpha}{\partial V_i}{\hphi} 
  = - \windon{\nu}{\partial V_i}{\Omega} \ensp,
\end{equation}
The right hand side is non-zero only if $\Omega$ is singular in $V_i$,
in which case the degree equals the instanton number of the gauge
singularities produced by $\Omega$ inside of $V_i$.  We have reduced
the problem to the calculation of the Hopf invariant of a single
monopole loop in a topologically trivial volume $V$.

\section{Generalized Hopf invariant}
\label{sec:gen-hopf}

In the modified axial gauge, it is possible to express the instanton
number in terms of monopole charges that can be calculated from
properties of the auxiliary Higgs field in the vicinity of the monopole
world-lines \cite{Jahn:1998nw}.  It would be desirable to establish a
similar relation in the general case.  Accordingly, we embed each
monopole loop into a loop of finite thickness and try to assign a Hopf
invariant to $\hphi$ on the surface $T$ of the thick loop.  This
surface is a higher-dimensional generalization of a tube and has the
topology of $\Sph^2\times\Sph^1$.  The coordinate corresponding to the
second factor can be interpreted as the proper time $\tau\in[0,2\pi]$
(in Euclidean space) of the monopole, the first factor as a sphere
surrounding the monopole at fixed $\tau$.  In quest of an invariant of
$\hphi\on_T$, we seek a characterization of the homotopy classes of
maps $\hphi\colon\Sph^2\times\Sph^1\to\Sph^2$.  These have been studied
in Ref.\ \cite{Taubes:1982ie}.  Following the ideas developed there, we
give a more explicit discussion that is better suited for our purposes.
A first characterization is given by the magnetic charge we have
introduced in the previous section.  It is the winding number of
$\hphi$ in its first argument for fixed $\tau$.  By continuity, it has
to be independent of $\tau$.  However, on a compact manifold the total
magnetic charge vanishes.  It is therefore not a good candidate for the
instanton number.

The most obvious ansatz for a further invariant, a naive generalization
of the Hopf invariant (\ref{eq:Hopf}), is only possible for $m=0$: the
magnetic charge is given by the integral of the pull-back $\omega_2$ of
the volume form on $\Sph^2$ for fixed $\tau$.  For $m\ne0$, it is
therefore not possible to write $\omega_2$ as a total differential.  In
this case, it is actually not possible to define an integer valued
invariant at all, since it turns out that the homotopy classes of maps
$\Sph^2\times\Sph^1\to\Sph^2$ with a given magnetic charge $m$ form the
group $\Z_{2|m|}$ rather than $\Z$ (as can be inferred from the
results of \cite{Taubes:1982ie}).

However, it is possible to generalize the Hopf invariant to a
restricted class of functions $\Sph^2\times\Sph^1\to\Sph^2$ with
magnetic charge $m\ne0$.  It is this invariant that will enable us to
establish a relation between instantons and monopoles in Sec.\ 
\ref{sec:loop}.  We consider maps $\phi:\Sph^2\times\Sph^1\to\Sph^2$
that map a fixed point on $\Sph^2$ to another fixed point $\phi_0$ on
the target $\Sph^2$ for every value of the second argument.  For
definiteness, we choose the first point to be the south pole.  In polar
coordinates $(\thet,\varphi)$ on $\Sph^2$, the restriction therefore
reads
\begin{equation}
  \label{eq:stos}
  \hphi(\thet{=}\pi,\varphi,\tau) = \phi_0 
\end{equation}
with $\phi_0\in\su(2)$ and $|\vect\phi_0|=1$.  (Recall that the target
$\Sph^2$ has been introduced as the unit sphere in $\su(2)$.)
Motivated by the relation (\ref{eq:hopf-vs-degree}) between the Hopf
invariant and the degree of a diagonalizing gauge transformation for
maps $\Sph^3\to\Sph^2$, we diagonalize $\hphi$,
\begin{equation}
  \label{eq:hphi-diag2}
  \hphi = \Omega^\dagg \sigma_3 \, \Omega \ensp,
\end{equation}
with $\Omega$ continuous on
$(\Sph^2\setminus\{\thet=\pi\})\times\Sph^1$.  For non-zero magnetic
charge $m$, $\Omega$ cannot be chosen continuous on all of
$\Sph^2\times\Sph^1$.  At the south pole, it has an Abelian discontinuity,
\begin{equation}
  \label{eq:Omega-disc}
  \Omega(\thet,\varphi,\tau) \to 
  \ee^{-\ii\psi(\varphi,\tau)\sigma_3} \,\Omega_0
  \itext{for}
  \thet \to \pi \ensp,
\end{equation}
related to the ambiguity of multiplying $\Omega$ with a diagonal matrix
from the left in Eq.\ (\ref{eq:hphi-diag2}).  $\Omega_0$ is a constant
matrix that diagonalizes $\phi_0$, i.e.,
$\phi_0=\Omega_0^\dagg\sigma_3\Omega_0$. 

Unfortunately, the analog of the degree for maps from
$\Sph^2\times\Sph^1$ to $\Sph^2$,
\begin{equation}
  \label{eq:nu-tube}
  \nu[\Omega] \defd \frac{1}{24\pi^2}
  \int_{\substack{\Sph^2\times\Sph^1\\\thet\ne\pi}} 
  \tr (\Omega^\dagg \dd\Omega)^3 \ensp,
\end{equation}
depends on the choice of the diagonalization matrix $\Omega$.  Under a
change $\Omega\to\omega\Omega$ with
$\omega=\ee^{\ii\chi(\thet,\varphi,\tau)\sigma_3}$, $\nu$ is not
invariant, because it is not additive for discontinuous $\Omega$,
\begin{equation}
  \label{eq:n-Omega-omega}
  \nu[\omega\Omega] = \nu[\Omega] + \nu[\omega]
  + \frac{1}{8\pi^2} \lim_{\eps\to0} \int\limits_{\thet=\pi-\eps}
  \tr \dd\Omega \, \Omega^\dagg \wedge \omega^\dagg \dd\omega \ensp.
\end{equation}
The winding number of the diagonal function $\omega$ vanishes, but the
surface term gives a contribution (on the boundary $\thet=\pi-\eps$, the
coordinate system $(\varphi,\tau)$ is right-handed since
$(\thet,\varphi,\tau)$ is right-handed on $\Sph^2\times\Sph^1$)
\begin{equation}
  \label{eq:n-surface}
  \begin{split}
    \frac{1}{8\pi^2} \lim_{\eps\to0} \smash{\int\limits_{\thet=\pi-\eps}}
    \tr \dd\Omega \, \Omega^\dagg \wedge \omega^\dagg \dd\omega
    &= \frac{1}{4\pi} \int \dd\psi(\varphi,\tau) \wedge
    \dd\chi(\pi,\varphi,\tau) \\
    &= \nab{\tau}[\psi] \, \nabt{\varphi}[\chi]
      - \nab{\varphi}[\psi] \, \nabt{\tau}[\chi] \ensp,
    \end{split}
\end{equation}
where we have introduced Abelian winding numbers, e.g.,
\begin{equation}
  \label{eq:Abelian-n}
  \nab{\varphi}[\psi] \defd \frac{1}{2\pi} \int_{0}^{2\pi}
  \dd\tau\, \frac{\partial\psi(\varphi_0,\tau)}{\partial\tau} \ensp.
\end{equation}
The other winding numbers are defined analogously.  They do not depend
on the values $\varphi_0$ respectively $\tau_0$.  The winding number
of $\chi$ for fixed $\tau$ vanishes since $\chi$ is continuous for
all $\thet<\pi$ including the north pole.  Hence,
\begin{equation}
  \label{eq:nu-prod}
  \nu[\omega\Omega] = \nu[\Omega] + \nab{\tau}[\psi] \nabt{\varphi}[\chi] \ensp.
\end{equation}
In the case at hand, the winding number of $\psi$ for fixed $\tau$ is
just the magnetic charge, $\nab{\tau}[\psi]=m$ (cf.\ Eq.\ 
(\ref{eq:m-psi})).  Since the discontinuous phase $\psi(\varphi,\tau)$
in Eq.\ (\ref{eq:Omega-disc}) changes by $-\chi(\pi,\varphi,\tau)$, it
is therefore possible to define an invariant as
\begin{equation}
  \label{eq:gen-hopf}
  \alpha_\varphi[\hphi] \defd -\nu[\Omega] - m \nab{\varphi}[\psi] \ensp.
\end{equation}
We will refer to this invariant as the \emph{generalized Hopf
  invariant} on $\Sph^2\times\Sph^1$.  It constitutes the desired
topological invariant for maps $\Sph^2\times\Sph^1\to\Sph^2$ with
magnetic charge $m$ that fulfill Eq.\ (\ref{eq:stos}).  It turns out
that (\ref{eq:gen-hopf}) is the only invariant and the homotopy classes
of such maps form the group $\Z$ \cite{Taubes:1982ie}.  The
restriction (\ref{eq:stos}) has increased the number of homotopy
classes since it restricts the set of possible deformations.  If
deformations that violate Eq.\ (\ref{eq:stos}) are allowed, maps with
$\alpha$ differing by multiples of $2 m$ can be deformed into each
other.  A mathematically more appealing definition of $\alpha$ is given
in Ref.\ \cite{Taubes:1982ie}.  It coincides with the more explicit
definition given here.

Note that the generalized Hopf invariant depends on the choice of the
coordinate $\varphi$, as indicated by the subscript on $\alpha$:
Consider, for instance, the coordinate system
$(\thet,\tilde\varphi,\tau)$ with $\tilde\varphi=\varphi+k\tau$ and
integer $k$ that is an admissible parameterization of
$\Sph^2\times\Sph^1$, too.  Under this change of coordinates,
$\nu[\Omega]$ is not altered, since the volume element occurring in the
integral (\ref{eq:nu-tube}) is invariant.  The winding number
(\ref{eq:Abelian-n}), however, changes,
\begin{equation}
  \label{eq:n-new}
  n_{\tilde\varphi=\varphi_0}[\psi] =
  \frac{1}{2\pi} \int_{0}^{2\pi}
  \dd\tau\, \frac{\dd\psi(\varphi_0-k\tau,\tau)}{\dd\tau} 
  = \nab{\varphi}[\psi] - k \nab{\tau}[\psi] \ensp,
\end{equation}
because the path $\{\varphi=\varphi_0-k\tau, \tau\in[0,2\pi]\}$ along
which the change of $\psi$ is calculated, is equivalent to the sum of
the original path $\{\varphi=\varphi_0,\tau\in[0,2\pi]\}$ and a path
that winds $k$ times around the negative $\varphi$-direction for fixed
$\tau$.  The generalized Hopf invariant therefore changes by $m^2 k$,
\begin{equation}
  \label{eq:alpha-tildephi}
  \alpha_{\tilde\varphi}[\hphi] = \alpha_{\varphi}[\hphi] + m^2 k \ensp.
\end{equation}

Furthermore, $\alpha_\varphi[\hphi]$ depends on the point on the
factor $\Sph^2$ of the domain (here the south pole) that is used to
formulate the constraint (\ref{eq:stos}).  One can show that a
different choice changes $\alpha_\varphi[\hphi]$ by
$2m\windon{\deg}{\Delta}{\hphi}$ with $\Delta=\gamma\times\Sph^1$
where $\gamma$ is a curve between the old and the new point.
To apply the above definition, one has to change the coordinate system
such that the new point corresponds to $\thet=\pi$, of course.

\begin{figure}[tbhp]
  \centering
  \epsfig{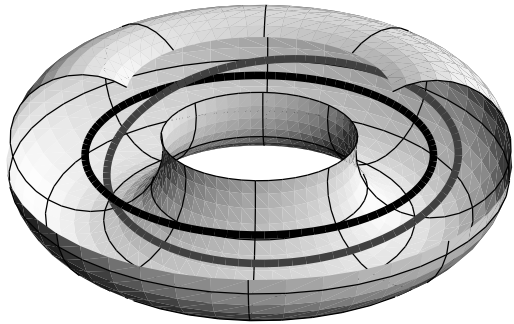}
  \caption{Generalized Hopf invariant as a linking number of preimages
    in a filled torus in three-space.  The picture shows an example
    with magnetic charge 1 (one preimage per point) and generalized
    Hopf invariant 1 (the preimages are linked once).}
  \label{fig:tlink}
\end{figure}

Geometrically, the generalized Hopf invariant is, as the original Hopf
invariant, given by the linking number of the preimages of two points
on the target $\Sph^2$ if we represent $\Sph^2\times\Sph^1$ as a filled
torus $\B^2\times\Sph^1$ in three-space with the boundary of the disk
$\B^2$ identified to one point -- the fixed point that is mapped to
$\phi_0$ in Eq.\ (\ref{eq:stos}) (cf.\ Fig.\ \ref{fig:tlink}).  The
ambiguity arising from different coordinates $\varphi$ is now replaced
by the ambiguity of different embeddings in three-space.  In order to
obtain the same definition as Eq.\ (\ref{eq:gen-hopf}), curves with
constant $\varphi$ on the surface of the filled torus must not `wind
around the torus', i.e., be topologically trivial in the complement of
the torus.  This fixes a possible `twist' of the torus.  Since, for a
charge $m$ configuration, each point has $m$ preimages and a twist
links every preimage with every other, it is obvious that it changes
the generalized Hopf invariant by $m^2$.  The example below will show
that the generalized Hopf invariant measures the twist of the Higgs
field.  In view of the relation between internal and real space present
in a field with non-zero winding number $m$, it is not surprising that
$\alpha_{\varphi}[\hphi]$ is also sensitive to a twist in real space.

\begin{figure}[tbhp]
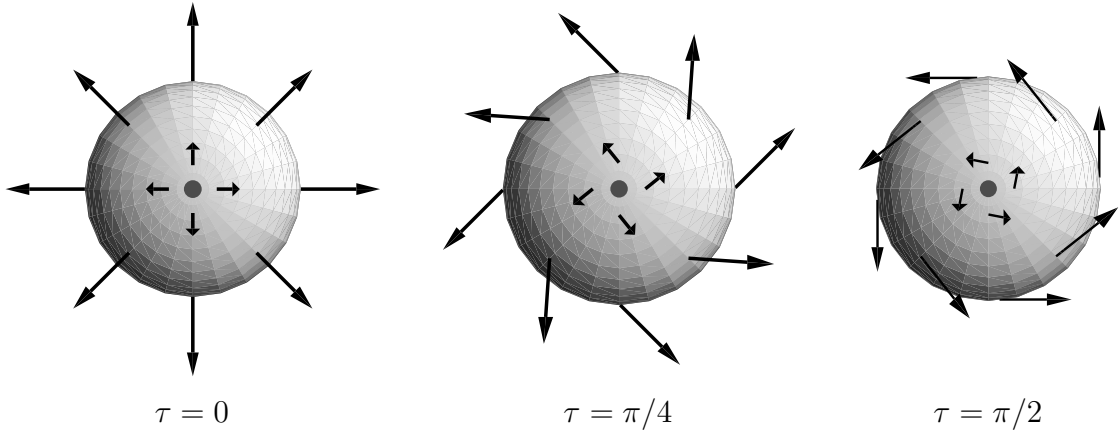

  \centering
  \makebox[\textwidth]{\hss
    $\begin{array}{ccccc}
      \vcenter{\hbox{\epsfig{file=twist0.epsi,width=.3325\textwidth}}} & \; &
      \vcenter{\hbox{\epsfig{file=twist1.epsi,width=.3135\textwidth}}} & \; &
      \vcenter{\hbox{\epsfig{file=twist2.epsi,width=.23275\textwidth}}} \\
      \tau=0 && \tau=\pi/4 && \tau=\pi/2 \vrule width0pt height1.5em
    \end{array}$\hss}
  \caption{Sketch of the Higgs field (\ref{eq:phi-m-k}) for $m=k=1$.
    The dot indicates the point that is mapped to $\phi_0$ as required
    by Eq.\ (\ref{eq:stos}).}
  \label{fig:twist}
\end{figure}

\paragraph{Example.}
As an example, consider the following auxiliary Higgs field with
magnetic winding number $m$,
\begin{equation}
  \label{eq:phi-m-k}
  \phi(\thet,\varphi,\tau) = 
  \begin{pmatrix}
    \sin\thet \cos (m\varphi-k\tau) \\
    \sin\thet \sin (m\varphi-k\tau) \\
    \cos\thet
  \end{pmatrix}
  \cdot \vect\sigma \ensp.
\end{equation}
It can be represented as a standard charge $m$ field on $\Sph^2$ that
is `twisted' around the three-axis along the world-line of the
monopole,
\begin{align}
  \label{eq:phi-twist}
  \phi(\thet,\varphi,\tau) &= \omega^\dagg(\tau) \phi(\thet,\varphi,0)
  \omega(\tau) \ensp, \\
  \omega(\tau) &= \ee^{-\ii k\tau\sigma_3/2} \ensp.
\end{align}
The field $\phi$ is displayed for some values of $\tau$ in Fig.\
\ref{fig:twist}.

Given a diagonalization at $\tau=0$,
\begin{equation}
  \label{eq:phi-m-k-diag}
  \phi(\thet,\varphi,0) = \Omega_1^\dagg(\thet,\varphi) \sigma_3
  \Omega_1(\thet,\varphi) \ensp,
\end{equation}
the $\tau$-dependent diagonalizing matrix can be represented as
\begin{equation}
  \Omega(\thet,\varphi,\tau) 
  = \omega^\dagg(\tau) \Omega_1(\thet,\varphi) \omega(\tau) \ensp.
\end{equation}
The factor $\omega^\dagg(\tau)$ is needed to make $\Omega$ periodic
also for odd $k$.  As argued above, the non-Abelian winding number of
$\Omega$ is the same as that of $\Omega_1$, because a shift of
$\varphi$ by a multiple of $\tau$ does not change it.  Since $\Omega_1$
depends on only two parameters, it vanishes,
$\nu[\Omega]=0$.  For $\thet\to\pi$, one finds
\begin{equation}
  \label{eq:Omega-m-k-lim}
  \Omega \to \ee^{\ii(k\tau-m\varphi)\sigma_3} \ii\sigma_2 \ensp,
\end{equation}
and therefore $\nab{\varphi}[\psi]=k$ and
\begin{equation}
  \label{eq:alpha-m-k}
  \alpha_{\varphi}[\hphi] = m k \ensp.
\end{equation}
We conclude that the generalized Hopf invariant is given by the product
of the magnetic charge and the number of times the Higgs field is
twisted along the monopole loop.  Obviously, the same is true for
twists of arbitrary configurations $\phi(\thet,\varphi,0)$.  It has
been observed that this kind of twist (called `Taubes winding' in Ref.\ 
\cite{Perez:1999ux}) gives rise to a non-vanishing instanton number
\cite{vanBaal:1997}.  This has been shown explicitly for a
finite-temperature instanton with non-trivial holonomy
\cite{Kraan:1998pm}.  The following sections investigate this relation
in detail for general configurations.

For unit charge monopoles, uniform twists give all possible values of
$\alpha_\varphi[\hphi]$.  For higher charges, there are additional
cases $0<\left|\alpha\right|<\left|m\right|$ that cannot be represented
in the simple form (\ref{eq:phi-twist}).  They correspond to fields
that are twisted only on a part of $\Sph^2$ that carries one unit (or
$m'<m$ units) of magnetic charge.

\section{Hopf invariant of a monopole loop}
\label{sec:loop}

We consider a single closed monopole loop $\calM$ where the Higgs field
vanishes (or is in the center for a group valued field).  Following the
strategy developed in Sec.\ \ref{sec:inst-genab}, we will embed the
monopole loop into a topologically trivial four-volume $V$.  By Eq.\ 
(\ref{eq:hopf-sum}), the contribution of the monopole loop to the instanton
number is then given by the Hopf invariant of $\hphi$ on the surface of
$V$.

As in the previous section, we first embed the monopole loop $\calM$
into a loop of finite thickness $r$,
\begin{equation}
  \label{eq:V_M}
  V_\calM \defd \bigl\{ x\in\R^4 \bigm\vert |x-y|\le r \text{ for some }
  y\in \calM \bigr\} \ensp.
\end{equation}
$r$ should be so small that $V_\calM$ does not become topologically
non-trivial by self-intersections.  Since we intend to apply the
definition of the generalized Hopf invariant given above, we choose an
isocurve $\calC$ of $\hphi$ on the surface $T$ of $V_\calM$,
\begin{equation}
  \label{eq:stos-C}
  \hphi \on_\calC = \hphi_0
  \itext{and} 
  \calC\subset T \defd \partial V_\calM 
  \ensp.
\end{equation}
A note on the existence of such a curve: On every section
$A\cong\Sph^2$ through $T$ and for every $\thet_0\in(0,\pi)$ there
exists a curve $\gamma$ on which $\hphi=(\thet_0,\varphi)$
and $\int_{\gamma}\dd\varphi\ne0$.  Let $A^-$
denote that connected component of $A\setminus\gamma$ where
the south pole is taken.  On moving $A$ along $\Sph^1$ through $T$,
$A^-$ changes continuously and cannot disappear because of
the non-vanishing integral.  The union of all $A^-$ gives an
open tube $T^-_{\thet_0}\subset T$.  On changing $\thet_0$,
$\gamma$, and therefore also $T^-_{\thet_0}$, can be chosen
to change continuously.  On the intersection
$T^-_{\pi}\defd\bigcap_{\thet_0\le\pi} T^-_{\thet_0}$, we have
$\hphi=\south$.  Since each section of $T^-_{\thet_0}$ along a $\Sph^2$
in $T$ is simply connected, this is also true for sections of
$T^-_{\pi}$.  Therefore, $T^-_{\pi}$ must contain a curve $\calC$ of
the required properties.

\begin{figure}[tbhp]
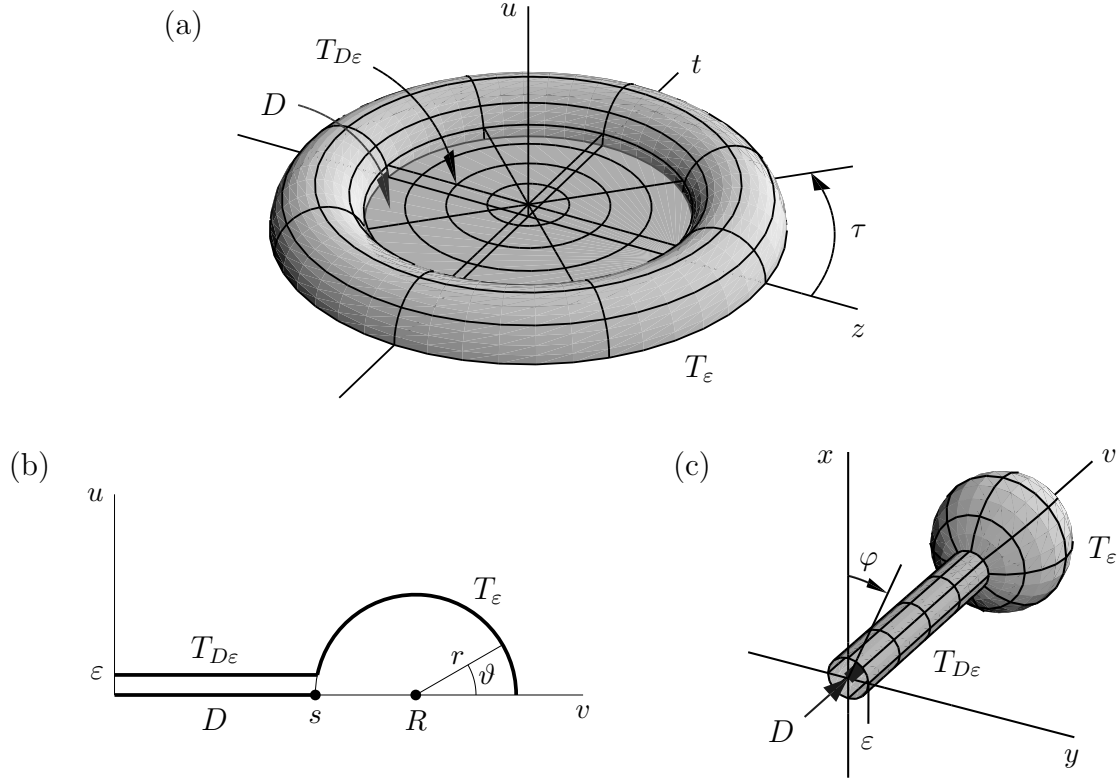

  \begin{minipage}[c]{\textwidth}
    \centering
    \makebox[\textwidth][c]{\vtop{\vbox{}\llap{(a)\quad}}%
    \vtop{\vbox{}\hbox{\psfrag{u}{\small$u$}%
        \psfrag{z}{\small$z$}%
        \psfrag{t}{\small$t$}%
        \psfrag{psi}{\small$\tau$}%
        \psfrag{Tep}{\raisebox{2mm}{\small\;\;\,\llap{$T_{D\varepsilon}$}}}%
        \psfrag{Te}{{\normalsize$T_{\varepsilon}$}}%
        \psfrag{D}{{\normalsize$\!D$}}%
        \epsfig{file=uzt-sh.epsi,width=.6\textwidth}}}}

    \vspace*{.5cm}
    
    \noindent\vtop{\vbox{}\hbox{(b)}}\quad\vtop{\vbox{}\hbox{%
    \psfrag{u}[r][Br]{{\small$u$}}%
    \psfrag{v}[t][B]{{\small$v$}}%
    \psfrag{theta}{\small$\vartheta$}%
    \psfrag{eps}[r][r]{\small$\varepsilon$}%
    \psfrag{s}[t][B]{\small$s$}%
    \psfrag{r}{\small$r$}%
    \psfrag{R}[t][B]{{\small$R$}}%
    \psfrag{D}[t][B]{{\normalsize$D$}}%
    \psfrag{Te}[bl][Bl]{{\normalsize$T_\eps$}}%
    \psfrag{Tep}[b][B]{{\normalsize$T_{D\eps}$}}%
    \setlength{\unitlength}{1mm}%
    \fmfframe(0,6)(0,0){\epsfig{file=uv-sh.epsi,width=0.45\textwidth}}}}
    \hfill
    \vtop{\vbox{}\hbox{(c)}}\quad\vtop{\vbox{}\hbox{\psfrag{v}{\small$v$}
    \psfrag{x}{\small$x$}
    \psfrag{y}{\small$y$}
    \psfrag{phi}{\small$\;\;\varphi$}
    \psfrag{eps}{\clap{\small$\eps\,$}}
    \psfrag{Tep}{\raisebox{-1mm}{\normalsize$\!T_{D\varepsilon}$}}
    \psfrag{Te}{{\normalsize$\!T_{\varepsilon}$}}
    \psfrag{D}{{\normalsize$\!D$}}
    \epsfig{file=xyv-sh.epsi,width=0.35\textwidth}}}
  \end{minipage}
  \caption{Manifolds used to define $V$:  (a) three-dimensional view
    for fixed $\varphi$, (b) two-dimensional view for fixed $\varphi$
    and $\tau$, (c) three-dimensional view for fixed $\tau$.}
  \label{fig:loop}
\end{figure}

Now, we close the loop with a two-dimensional sheet $D$ (reminiscent of
a Dirac sheet) that has $\calC$ as its boundary,
\begin{equation}
  \label{eq:D-C}
  \partial D = \calC \ensp,
\end{equation}
and intersects $V_\calM$ only there.  For $r\to0$, the condition
(\ref{eq:stos-C}) can be represented in terms of $D$: it requires that
$D$ emerges from $\calM$ in a direction where $\hphi=\hphi_0$.  We
complement $V_\calM$ by a sheet of finite thickness $\eps<r$ around
$D$,
\begin{equation}
  V_D \defd \bigl\{ x\in\R^4 \bigm\vert |x-y|\le\eps \text{ for some }
  y\in D \bigr\} \cong \B^2\times\B^2 \ensp,
\end{equation}
to define the topologically trivial volume $V$,
\begin{equation}
  \label{eq:V}
  V = V_\calM \cup V_D \ensp.
\end{equation}
Eventually, we will perform the limit $\eps\to0$.  
We decompose the surface of $V$ into parts around the loop and the
sheet,
\begin{gather}
  \partial V = T_\eps \cup T_{D\eps} \ensp, 
  \\
  T_{\eps} \defd
  \overline{T\setminus V_D} 
  \cong \B^2\times\Sph^1 \ensp, 
  \\
  T_{D\eps} \defd 
  \overline{\partial V_D \setminus V_\calM} 
  \cong \Sph^1\times\B^2 \ensp.
\end{gather}
The various manifolds are sketched in Fig.\ \ref{fig:loop} for the
example of a loop in the $z$-$t$-plane, $x=y=0$, $z^2+t^2=R^2$, using
double polar coordinates in space-time,
\begin{equation}
  \label{eq:polar-conformal}
  x + \ii y = u \ee^{\ii\varphi}
  \itext{and}
  z + \ii t = v \ee^{\ii\tau} \ensp.
\end{equation}
The tube $T$ can be parameterized by the coordinates
$\thet=\arctan(u/(v-R))$, $\varphi$ and $\tau$ that have the same
orientation as in Sec.\ \ref{sec:gen-hopf}.  A double set of polar
coordinates $(u,\varphi,v,\tau)$ with $u=0$ on $D$ and
$v=v_0=\text{const}$ on $\calC$ can be chosen for any $\calM$ and $D$
and will be used in the following.

The intersection
\begin{equation}
  t \defd T_{\eps} \cap T_{D\eps} = \partial T_{\eps}
  = \partial T_{D\eps} 
  \cong \Sph^1 \times \Sph^1 
\end{equation}
is parameterized by the coordinates $\varphi$ and $\tau$.

\null

Since there is no local representation of the Hopf invariant, we can
not calculate separate contributions from $T_\eps$ and $T_{D\eps}$ to
$\windon{\alpha}{\partial V}{\hphi}$.  Therefore, we diagonalize
$\hphi$ on $\partial V\cong\Sph^3$,
\begin{equation}
  \label{eq:phi-diag-TT}
  \hphi = \Omega^\dagg \sigma_3 \Omega
  \itext{on}
  \partial V \ensp,
\end{equation}
and calculate the contributions to $\nu[\Omega]$, which is by Eq.\ 
(\ref{eq:hopf-inst-indiv}) equal to the negative of the desired Hopf
invariant.

In the limit $\eps\to 0$, the intersection $t$ reduces to the curve
$\calC$.  Condition (\ref{eq:stos-C}) implies that, in this limit,
$\Omega$ is constant up to a diagonal factor,
\begin{equation}
  \label{eq:Omega-diag-t}
  \Omega \to \ee^{-\ii\psi(\varphi,\tau) \sigma_3} \Omega_0
  \itext{on}
  t
  \itext{for}
  \eps \to 0 \ensp.
\end{equation}
As in Sec.\ \ref{sec:gen-hopf}, the winding number of $\psi$ for fixed
$\tau$ gives the magnetic charge of the monopole singularity,
\begin{equation}
  \label{eq:m_psi}
  \nab{\tau}[\psi] = m \ensp.
\end{equation}
The interpretation of the winding number for fixed $\varphi$ can be
found by noting that $T_{D\eps}$ approaches the sheet $D$ in the limit
$\eps\to0$, and therefore
\begin{equation}
  \label{eq:Omega-TD}
  \Omega \to \ee^{\ii\chi(\varphi,v,\tau)\sigma_3}
  \tilde\Omega(v,\tau)
  \itext{on} T_{D\eps} \itext{for} \eps\to0 \ensp,
\end{equation}
where $\tilde\Omega$ is independent of $\varphi$ and diagonalizes
$\hphi$ on $D$,
\begin{equation}
  \label{eq:phi-diag-D}
  \hphi = \tilde\Omega^\dagg \sigma_3 \tilde\Omega
  \itext{on} D \ensp.
\end{equation}
On the boundary $\partial D=\calC$, also $\tilde\Omega$ is constant up
to a diagonal factor,
\begin{equation}
  \label{eq:Omega-diag-C}
  \tilde\Omega = \ee^{-\ii\tilde\psi(\tau) \sigma_3} \Omega_0
  \itext{on}
  \calC \ensp.
\end{equation}
By the same way as the winding number of $\psi$ is related to the
magnetic charge, that of $\tilde\psi$ is related to the degree of
$\hphi$ on $D$ (cf.\ Eqs.\ (\ref{eq:m-deg-phi}) and (\ref{eq:m-psi})),
\begin{equation}
  \label{eq:deg-D}
  n[\tilde\psi] = \deghphiD \ensp.
\end{equation}
This degree is well-defined since the boundary $\calC$ of $D$ is mapped
to a single point.  $D$ is therefore effectively compactified to
$\Sph^2$.  It can be interpreted as the flux through $D$ produced by
the gauge fixing transformation.  However, since the flux stems from a
finite magnetic field, unlike the flux of the monopole singularity, it
cannot be distinguished from the flux already present before gauge
fixing.

Furthermore, since the two expressions (\ref{eq:Omega-diag-t}) and
(\ref{eq:Omega-TD}) for $\Omega$ on $t=\partial T_{D\eps}$ for
$\eps\to0$ have to coincide, the relation
$\psi(\varphi,\tau)=\tilde\psi(\tau)-\chi(\varphi,v_0,\tau)$ follows
($v=v_0$ on $\calC$).  $\chi$ is continuous also for $v\to0$.
Therefore, its winding number with respect to $\tau$ vanishes and the
corresponding winding numbers of $\psi$ and $\tilde\psi$ are identical,
whence
\begin{equation}
  \label{eq:n-phi-psi}
  \nab{\varphi}[\psi] = \deghphiD \ensp.
\end{equation}
The winding number with respect to $\varphi$ is the negative of that of 
$\psi$,
\begin{equation}
  \label{eq:n-tau-chi}
  \nab{\tau}[\chi] = -m \ensp.
\end{equation}

\null

We can now express the contributions to $\windon{\nu}{\partial
  V}{\Omega}$ in the limit $\eps\to0$ in terms of $\hphi$.  For
$T_\eps$, we insert the winding number of $\psi$ into the definition
of the generalized Hopf invariant, Eq.\ (\ref{eq:gen-hopf}), to obtain
\begin{equation}
  \label{eq:hopf-nu-T}
  \windon{\alpha_D}{T}{\hphi} = 
  - \lim_{\eps\to0} \windon{\nu}{T_\eps}{\Omega} - m \deghphiD
  \ensp.
\end{equation}
We have replaced the subscript $\varphi$ on $\alpha$ by $D$, because
$D$ determines the coordinate $\varphi$ up to homotopy:  $\varphi$ is
that angle on the torus $t$ that can be continuously extended to the
whole tube $T_{D\eps}$ around $D$.  Obviously, this is not the case for 
$\tau$ ruling out an admixture of $\tau$ to $\varphi$.

\begin{figure}[tbhp]
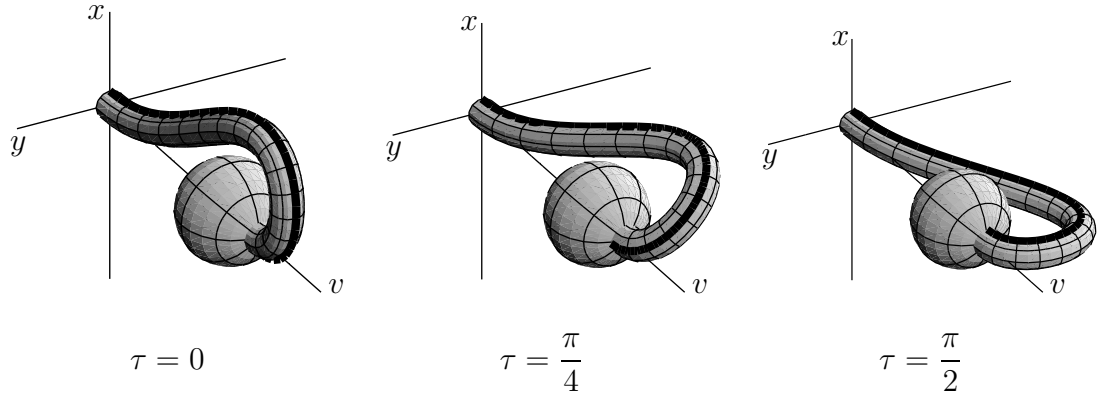

  \begin{minipage}[c]{\textwidth}
    \centering
    {\psfrag{v}{$v$}
      \psfrag{x}{$x$}
      \psfrag{y}{$y$}
      \psfrag{psi}{$\tau_0$}
      \epsfig{file=ddep0.epsi,width=.3\textwidth}%
      }%
    \hspace{.5cm}%
    {%
      \psfrag{v}{$v$}
      \psfrag{x}{$x$}
      \psfrag{y}{$y$}
      \psfrag{psi}{$\tau_0$}
      \epsfig{file=ddep1.epsi,width=.3\textwidth}%
      }%
    \hspace{.5cm}%
    {%
      \psfrag{v}{$v$}
      \psfrag{x}{$x$}
      \psfrag{y}{$y$}
      \psfrag{psi}{$\tau_0$}
      \epsfig{file=ddep2.epsi,width=.3\textwidth}%
      }%
    
    \vspace*{.5cm}
    
    \makebox[\textwidth]{\hfill $\tau=0$ \hfill
      \hfill $\displaystyle\tau=\frac{\pi}{4}$ \hfill
      \hfill $\displaystyle\tau=\frac{\pi}{2}$ \hfill}
  \end{minipage}
  \caption{Alternative choice of the sheet $D$.  For details see text.}
  \label{fig:ddep}
\end{figure}

For the second contribution, we note that, since
$T_{D\eps}\cong\Sph^1\times\B^2$ has the same topology as $T_\eps$, we
can apply the relation (\ref{eq:nu-prod}) for the non-Abelian winding
number of a product to Eq.\ (\ref{eq:Omega-TD}).  The angles $\tau$ and
$\varphi$ have exchanged their roles:
\begin{equation}
  \label{eq:nu-TD}
    \lim_{\eps\to0} \windon{\nu}{T_{D\eps}}{\Omega} =
    \nu[\tilde\Omega] +
    \nab{\varphi}[\tilde\psi] \nab{\tau}^{v=v_0}[\chi]
    = - \deghphiD \, m \ensp,
\end{equation}
where we have used the fact that $\nu[\tilde\Omega]$ vanishes since
$\tilde\Omega$ depends on only two parameters and have inserted the
winding numbers (\ref{eq:deg-D}) and (\ref{eq:n-tau-chi}).  Putting
Eqs.\ (\ref{eq:hopf-nu-T}) and (\ref{eq:nu-TD}) together, we obtain
\begin{equation}
  \label{eq:nu-del-V}
  \lim_{\eps\to0} \windon{\nu}{\partial V}{\Omega} =
  \lim_{\eps\to0} \left(
    \windon{\nu}{T_\eps}{\Omega} 
    \windon{\nu}{T_{D\eps}}{\Omega}
  \right) =
  - \windon{\alpha_D}{T}{\hphi} - 2 m \deghphiD \ensp.
\end{equation}
The Hopf invariant of $\hphi$ on $\partial V$ is therefore given by
\begin{equation}
  \label{eq:alpha-loop}
  \windon{\alpha}{\partial V}{\hphi}
  = \alphaTD + 2 m \deghphiD \ensp.
\end{equation}
This is the desired expression for the contribution of a monopole loop
to the instanton number (cf.\ Eq.\ (\ref{eq:hopf-sum})).  While
$\alphaTD$ depends on the position of the sheet $D$, it is independent
of the values of $\hphi$ on $D$, as indicated.  The latter enter
through the term $\deghphiD$, though.  The instanton number is
therefore not given by properties of the auxiliary Higgs field near the
monopole singularity only.  The instanton number modulo $2m$, however,
is:
\begin{equation}
  \label{eq:nu-mod-2m}
  \windon{\alpha}{\partial V}{\hphi} = \alphaTD \pmod{2m} \ensp.
\end{equation}
One can show that also the dependence on the position of $D$ disappears
here, since a different choice of $D$ changes $\alphaTD$ only by
multiples of $2m$:  We have already seen in Sec.\ \ref{sec:gen-hopf}
that a change of the curve $\calC$ used for the condition
(\ref{eq:stos-C}) generates such a shift.  There, however, a change of
the coordinate $\varphi$ produced a shift by $m^2$.  Here, only a shift
by $2m^2$ is possible.  The reason is that the embedding of
$\Sph^2\times\Sph^1$ into $\R^4$ given by $T$ fixes the coordinate
$\varphi$ up to multiples of $2\tau$ (up to homotopy).  Figure
\ref{fig:ddep}, for instance, shows an alternative choice of the sheet
$D$ for the loop of Fig.\ \ref{fig:loop}.  Consider first a sheet that
stays at the position indicated in the first picture for all $\tau$.  A
curve of constant $\varphi$ corresponds to a $\tau$-independent point
on the circle where the tube meets the sphere.  Now consider a sheet
that winds once around the sphere while $\tau$ changes by $2\pi$ as
indicated in the other pictures.  Since $\varphi$ has to be continuous
on $T_{D\eps}$, a curve of constant $\varphi$ has to be homotopically
equivalent to a $\tau$-independent point in the $v=0$-plane.  This is
indicated by the thick lines on the tubes for a point on the positive
$x$-axis.  On the intersection of tube and sphere, the curve of
constant $\varphi$ now winds twice around the circle as $\tau$ changes
by $2\pi$.  This sheet therefore corresponds to a new coordinate
$\tilde\varphi=\varphi+2\tau$.  Obviously a shift by only $\tau$ is not
possible.

Consequently, we can assign a unique $\Z_{2|m|}$-valued generalized
Hopf invariant to $\hphi\on_{T_i}$,
\begin{equation}
  \label{eq:alpha-Z2m}
  \windon{\alpha}{T}{\hphi} \equiv \alphaTD \bmod 2 m \in \Z_{2|m|}
  \ensp,
\end{equation}
and write
\begin{equation}
  \label{eq:alpha-noD}
  \windon{\alpha}{\partial V}{\hphi} \bmod 2 m =
  \windon{\alpha}{T}{\hphi} \ensp.
\end{equation}
Since the group of homotopy classes of maps from $\Sph^2\times\Sph^1$
to $\Sph^2$ with magnetic winding number $m$ is $\Z_{2|m|}$ (cf.\ Sec.\ 
\ref{sec:gen-hopf}), this is the maximal information that can be
expected.

One could argue that it is possible to get rid of the additional term
in Eq.\ (\ref{eq:alpha-loop}) by choosing a sheet $D$ on which $\hphi$
is constant, $\hphi\on_D=\hphi_0$.  However, this is not possible in
general.  If the Hopf invariant $\alpha[\hphi\on_{\partial V}]$ is
non-zero, $\hphi$ takes all possible values on $\partial V$.  This
implies that the preimages of all points extend to the exterior of $V$
(and some even to infinity if the total instanton number is non-zero).
One therefore has to expect that also an isosurface $D$ whose boundary
is a monopole loop leaves $V$.  Such a $D$ cannot be used to
identify the contribution of an individual monopole loop to the
instanton number in the way described here.

\null

\begin{figure}[tbhp]
  \centering
  \epsfig{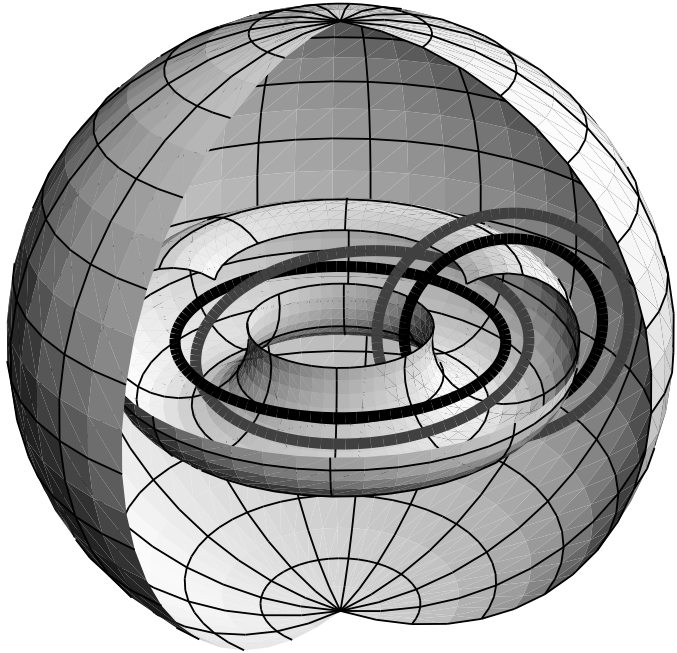}
  \caption{Decomposition $\Sph^3=\B^2\times\Sph^1\cup\Sph^1\times\B^2$
    and representation of $\windon{\alpha}{\partial V}{\hphi}$ as a
    linking number of preimages.  $\Sph^3$ is represented as a three-ball
    with its surface identified to a point.}
  \label{fig:s3tori}
\end{figure}

The result (\ref{eq:alpha-loop}) can also be understood geometrically:
The decomposition $\partial V=T_\eps\cup T_{D\eps}$ corresponds
topologically to the decomposition of $\Sph^3$ into two filled tori,
$\Sph^3=\B^2\times\Sph^1\cup\Sph^1\times\B^2$ (cf.\ Fig.\ 
\ref{fig:s3tori}).  The Hopf invariant of $\hphi$ on $\partial V$ is
given by the linking number of the preimages of two points.  Each point
has $m=\windon{\deg}{T}{\hphi}$ preimages in the filled torus
corresponding to $T_\eps$ and $\deghphiD$ preimages in the one
corresponding to $T_{D\eps}$ if the orientation of the preimages is
taken into account.  If we furthermore choose the decomposition into
the two tori compatible with the coordinate $\varphi$ around $\calC$ in
the same way as the embedding of the torus into $\R^3$ in Sec.\ 
\ref{sec:gen-hopf}, the linking number of the preimages in $T_\eps$ is
given by $\alphaTD$.  The preimages in $T_{D\eps}$ do not link since
$\hphi$ becomes $\varphi$-independent in the limit $\eps\to0$.
Finally, we have to take into account the linking between the preimages
in $T_\eps$ and $T_{D\eps}$.  This gives the remaining term
$2m\deghphiD$ in Eq.\ (\ref{eq:alpha-loop}).

\begin{figure}[tbhp]
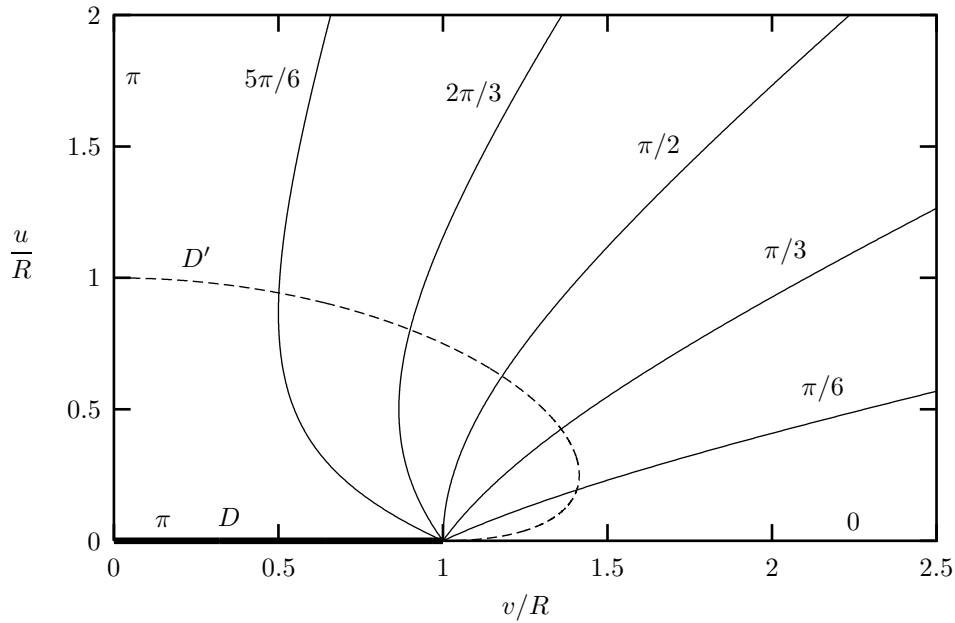

  \centering
  {\fontsize{10}{12}\selectfont \input beta }
  \vspace*{-0.5cm}
  \caption{Contour plot of the polar angle parameter $\beta(u,v)$.  Two 
    alternative choices ($D$ and $D'$) of the Dirac sheet are included.}
  \label{fig:beta}
\end{figure}

\paragraph{Monopole loop for instanton solution.}
The authors of \cite{Brower:1997js} have found solutions to the
differential maximal Abelian gauge condition for the single instanton
solution \cite{Belavin:1975,Jackiw:1977fs} that correspond to closed
monopole loops of various radii.  Although the global minimum of the
gauge fixing functional (\ref{eq:RmAg}) is only reached in the limit of
zero radius, it is conjectured that a small perturbation from, e.g., a
nearby instanton can stabilize a finite radius.  To check our
result,\footnote{In the case of the maximal Abelian gauge, it is also
  valid for the space-time $\R^4$, because the finiteness of the gauge
  fixing functional \ref{eq:RmAg} guarantees the validity of Eq.\ 
  \ref{eq:pure-gauge-higgs}.}  we calculate the contributions to Eq.\ 
(\ref{eq:alpha-loop}) for the explicit solution that has been given in
\cite{Brower:1997js} for the limit in which the radius of the monopole
loop is much smaller than the radius of the instanton.  In the double
polar coordinates of Eq.\ (\ref{eq:polar-conformal}) in space-time and
in spherical polar coordinates in target space, the solution for a
regular gauge instanton reads
\begin{equation}
  \label{eq:phi-inst}
  \phi = 
  \begin{pmatrix}
    \sin\beta \cos(\varphi+\tau) \\
    \sin\beta \sin(\varphi+\tau) \\
    \cos\beta
  \end{pmatrix}
  \cdot \vect\sigma \ensp,
\end{equation}
where $\beta$ is a function of $u$ and $v$ only,
\begin{equation}
  \label{eq:beta}
  \beta(u,v) = \thet_+ + \thet_- 
  \itext{where}
  \tan\thet_\pm = \frac{u}{v\pm R} \ensp.
\end{equation}
The angles $\thet_\pm$ can be chosen continuous modulo $2\pi$
everywhere with the exception of the circle $u=0$, $v=R$, where the
monopole singularity arises (cf.\ Fig.\ \ref{fig:thetas} copied from
\cite{Brower:1997js}).  A contour plot of $\beta(u,v)$ is shown in
Fig.\ \ref{fig:beta}.
\begin{figure}[tbhp]
  \centering
  {\fontsize{10}{12}\selectfont \input thetas }
  \caption{Variables $\thet_+$ and $\thet_-$}
  \label{fig:thetas}
\end{figure}
Since $\beta$ tends to $0$ or $\pi$ for $u\to0$ or $v\to0$, there are
no additional singularities due to the angles $\varphi$ and $\tau$.
For $u^2+v^2\to\infty$, $\phi$ tends to the standard Hopf map
\cite{Nakahara:1990} with $\tau$ substituted by $-\tau$ and therefore
carries a Hopf invariant of $-1$.  A gauge transformation that
diagonalizes $\phi$ removes the instanton winding number from the gauge
potential at infinity and produces a gauge singularity along the
monopole loop (and on a sheet $D$) that carries the same winding
number.

In order to calculate the contributions to Eq.\ (\ref{eq:alpha-loop}),
we note that near the monopole $\thet_+\to0$ and $\thet_-$ complements
$\varphi$ to a set of spherical polar coordinates on the sphere around
the monopole.  Finally, $\tau$ measures the position along the monopole
loop.  A natural choice for the sheet $D$ is $u=0$, $v\le R$ as in
Fig.\ \ref{fig:loop} where $\thet_-=\pi$ and $\phi=-\sigma_3$.  The
condition (\ref{eq:stos-C}) is therefore fulfilled for every tube
around the monopole loop.  Furthermore, the coordinate $\varphi$ is
compatible with the sheet $D$ since it can be defined globally on a
tube $u=\eps$, $v\le R$ around the sheet $D$.  Since $\beta\sim\thet_-$
near the monopole loop, $\phi$ is identical to the field
(\ref{eq:phi-m-k}) from the example in Sec.\ \ref{sec:gen-hopf} for
$m=k=1$ and the first term in Eq.\ (\ref{eq:alpha-loop}) is therefore
\begin{equation}
  \windon{\alpha_D}{T}{\phi}=-1 \ensp.
\end{equation}
On the sheet $D$, $\phi$ is constant.  The corresponding degree
therefore vanishes,
\begin{equation}
  \windon{\deg}{D}{\hphi} = 0 \ensp,
\end{equation}
and the second term in (\ref{eq:alpha-loop}) does not contribute.  We
obtain the expected result $\nu=1$.  Note that this is a non-generic
case where the argument of the paragraph after Eq.\ 
(\ref{eq:alpha-noD}) is circumvented in a special way: while indeed the
preimages of all points extend to infinity, that of $\beta=\pi$ splits
at the origin into the plane $v=0$ and the sheet $D$ due to a vanishing
Jacobi matrix of $\phi$.

To see how the contributions to the instanton number depend on the
sheet $D$ chosen, we repeat the calculation for an alternative sheet
$D'$ indicated schematically in Fig.\ \ref{fig:beta}.  The indicated
relation between $u$ and $v$ is complemented by the condition
$\varphi=\varphi_0$.  The angle $\psi$ is not constrained.  By using
the right-handed set of coordinates
$(\thet'=\pi-\thet_-,\varphi'=-\varphi,\tau)$ and the formulae in Sec.\ 
\ref{sec:gen-hopf}, one finds that in this case
\begin{equation}
  \windon{\alpha_{D'}}{T}{\phi}=+1 \ensp.
\end{equation}
The Higgs field on the sheet $D'$ is not constant any more but takes
all values on $\Sph^2$ as can be seen in Fig.\ \ref{fig:beta},
\begin{equation}
  \label{eq:phi-D'}
  \phi\on_{D'} = 
  \begin{pmatrix}
    \sin(\pi-\rho) \cos(\varphi_0+\tau) \\
    \sin(\pi-\rho) \sin(\varphi_0+\tau) \\
    \cos(\pi-\rho)
  \end{pmatrix}
  \ensp,
\end{equation}
where $\rho$ is a suitable radial coordinate with range $[0,\pi]$ on
$D'$.  Due to the occurrence of $\pi-\rho$, this map has degree
\begin{equation}
  \windon{\deg}{D'}{\phi}=-1 \ensp.  
\end{equation}
The magnetic charge is still $+1$ because we have not changed the
orientation of $\tau$.  The Hopf invariant on the surface $\partial V'$
around $T$ and $D'$ is therefore again $\windon{\alpha}{\partial
  V'}{\phi}=1-2=-1$.  The contributions from the generalized Hopf
invariant and the Higgs field on the sheet $D$, however, have changed.

By `twisting' the sheet $D'$, i.e., replacing the condition
$\varphi=\varphi_0$ by $\varphi=\varphi_0+n\tau$, one can obtain any
odd value $\windon{\alpha_D'}{T}{\phi}=2n+1$ and the appropriate value
$\windon{\deg}{D}{\phi}=-n-1$ that yield a total
$\windon{\alpha}{\partial V}{\phi}=-1$.

\section{Topologically non-trivial monopole loops}
\label{sec:nontriv}

The procedure of closing the individual loops by  sheets cannot
be applied to loops that are topologically non-trivial in space-time.
The simplest geometry where this can occur is $\Sph^3\times\Sph^1$.
Topologically non-trivial loops wind around the second factor. 
For simplicity, we assume all fields to be periodic in the second
factor.  This can always be accomplished by a gauge transformation.  We 
map $\Sph^3$ by stereographic projection to $\R^3$ such that that
there is no monopole at the point that is mapped to infinity.  In this
case, the fields tend to a pure gauge at infinity,
\begin{equation}
  \label{eq:pure-gauge-S2S1}
  \begin{aligned}
    A(x) & \sim U^\dagg(\hx,t) \,\dd U(\hx,t) \\
    \phi(x) & \to U^\dagg(\hx,t) \,\phi_0 \, U(\hx,t)
    \equiv \phi_\infty(\hx,t) 
  \end{aligned}
  \itext{for} |\vect x| \to \infty \ensp,
\end{equation}
and the instanton number is given by the winding number of the map
$U\colon\Sph^2\times\Sph^1\to\SU(2)$ which can be expressed as an
integral over the same density as for maps $\Sph^3\to\SU(2)$,
\begin{equation}
  \label{eq:nu-U-S2S1}
  \nu = \nu[U] 
  \defd \frac{1}{24\pi^2} \int_{\Sph^2\times\Sph^1} \!\tr \left[
    (U^\dagg \dd U)^3
  \right] \ensp.
\end{equation}
Since the total magnetic charge on the compact manifold $\Sph^3$
necessarily vanishes, $\hphi_\infty\colon\Sph^2\times\Sph^1\to\Sph^2$
has magnetic winding number zero.  As already mentioned in Sec.\ 
\ref{sec:gen-hopf}, the set of homotopy classes of such maps is $\Z$
and is parameterized by a Hopf invariant defined in an analogous way as
for maps $\Sph^3\to\Sph^2$.  Consequently, also the relation between
the winding number of $U$ and the Hopf invariant of $\hphi_\infty$
remains the same,
\begin{equation}
  \label{eq:nu-Hopf-S2S1}
  \nu[U] = - \alpha[\hphi_\infty] \ensp.
\end{equation}

However, the procedure advocated in Sec.\ \ref{sec:inst-genab} is not
directly applicable here because it is not possible to embed
topologically non-trivial monopole loops into topologically trivial
volumes.  If we embed a single topologically non-trivial loop into a
topologically non-trivial volume $V$, the auxiliary Higgs field has a
non-zero magnetic winding number on $\partial V$.  It is therefore not
possible to assign a unique Hopf invariant to it.  The best we can do
in order to decompose the Hopf invariant, is to group the monopole
loops into neutral sets and embed each set into a volume that is
topologically as simple as possible, i.e., equivalent to
$\B^3\times\Sph^1$.  If this is done, the Hopf invariant of
$\hphi_\infty$ again splits into contributions from the boundaries of
the volumes $V_i$,
\begin{equation}
  \label{eq:alpha-Vi}
  -\nu = \alpha[\hphi_\infty] 
  = \sum_i \windon{\alpha}{\partial V_i}{\hphi}
  \ensp,
\end{equation}
where topologically trivial loops are treated as before.

\begin{figure}[tbhp]
  \centering
  {\small
    \psfrag{T1}[tr][tr]{{\normalsize$T_{1\varepsilon}\!\!$}}%
    \psfrag{TD1}[b][b]{{\normalsize$T^1_{\varepsilon}$}}%
    \psfrag{T2}[tl][tl]{{\normalsize$T_{2\varepsilon}$}}%
    \psfrag{TD2}[b][b]{{\normalsize$T^2_{\varepsilon}$}}%
    \psfrag{T3}[tl][tl]{{\normalsize$T_{3\varepsilon}$}}%
    \psfrag{phi}[t][t]{{\small$\varphi$}}%
    \psfrag{s}[t][t]{{\small$s$}}%
    \epsfig{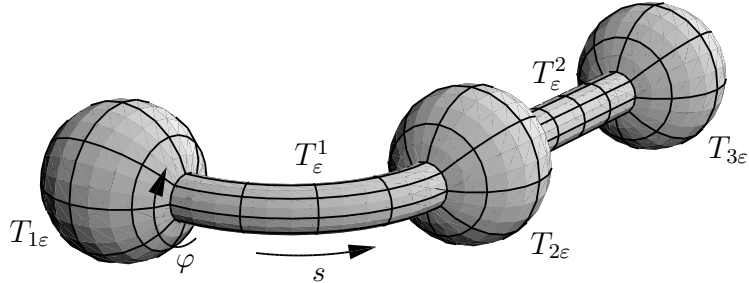}}
  \caption{Three-dimensional section of space-time for fixed $\tau$
    with tubes around monopoles and sheets.}
  \label{fig:D-nontriv}
\end{figure}

To complete the calculation of the instanton number, we only have to
consider a single set of $N$ topologically non-trivial monopole loops
$\calM_i$ with magnetic charges $m_i$ and total charge zero,
\begin{equation}
  \label{eq:sum-mi}
  \sum_{i=1}^N m_i = 0 \ensp.
\end{equation}
In order to construct the volume $V\cong\B^3\times\Sph^1$ around these,
we first embed the individual loops $\calM_i$ into thick loops
$V_{\calM_i}$ with boundaries $T_i\cong\Sph^2\times\Sph^1$.
Then we connect the `tubes' $T_i$ by $N-1$ sheets $D^{\beta}$ that do
not intersect with each other and intersect with the tubes on curves
$\calC_i^\beta$ where $\hphi$ is constant (cf.\ Fig.\ 
\ref{fig:D-nontriv}),
\begin{gather}
  \begin{aligned}
    D^\beta \cap D^\gamma &= \emptyset 
    \itext{for}\beta\ne\gamma \ensp, \\
    D^\beta \cap T_i &= \calC_i^\beta \subset \partial D^\beta \ensp, 
  \end{aligned} \\
  \hphi\on_{\calC_i^\beta} = \phi_i^\beta = \text{const} \ensp.
\end{gather}
We assume that two of the tubes ($T_1$ and $T_N$) intersect only with
one and the others with two  sheets.  This means that
monopoles and  sheets form an open chain.  The tubes and sheets will be 
numbered consecutively.

As for the case of topologically trivial monopole loops, we introduce
thick  sheets $V_{D^\beta}\cong\B^2\times I\times\Sph^1$ and
decompose the surface around the union $V$ of all thick loops and
sheets,
\begin{equation}
  \label{eq:V-nontriv}
  V \equiv \union_i V_{\calM_i} \cup \union_\beta V_{D^\beta} \ensp,
\end{equation}
into parts around loops and sheets,
\begin{align}
  \partial V &= \union_i T_{i\eps} \cup
  \union_\beta T^\beta_\eps  \ensp, \\
  T_{i\eps} &\defd 
  \overline{T_i \setminus \union_\beta V_{D^\beta}} \ensp, \\
  T^\beta_\eps &\defd 
  \overline{\partial V_{D^\beta} \setminus \union_i V_{\calM_i}}
  \ensp.
\end{align}
The topology of these manifolds is as follows,
\begin{align}
  \partial V &\cong \Sph^2\times\Sph^1 \ensp, \\
  T_{i\eps} &\cong \begin{cases}
    \B^2\times\Sph^1 &\text{for\quad}i=1,N \ensp, \\
    \Sph^1\times I \times\Sph^1 &\text{for\quad}i=2,\ldots,N-1 \ensp,
  \end{cases} \\
  \label{eq:T-beta-eps}
  T^\beta_\eps &\cong \Sph^1\times I \times\Sph^1 \ensp.
\end{align}
The intersections still have the topology of tori,
\begin{equation}
  \label{eq:t-nontriv}
  t_i^\beta \equiv T_{i\eps} \cap T^\beta_\eps
  \cong \Sph^1 \times \Sph^1 \ensp.
\end{equation}
We assume $\eps$ to be so small that the thick sheets $V_{D^\beta}$ do
not intersect.  $t_i^\beta$ will be parameterized by two angles
$\varphi$ and $\tau$, where $\tau$ runs along the monopole loops and
can be defined globally on $\partial V$ while $\varphi$ measures the
angle around the sheets.  It can be defined globally on $\partial V$
with the exception of one point on both $T_{1\eps}$ and $T_{N\eps}$.
We complement $\varphi$ and $\tau$ with a third coordinate $s$ such
that $s$ and $\varphi$ are spherical polar coordinates on the factor
$\Sph^2$ of $\partial V$ and $s=s_i^\beta$ is constant along the curves
$\calC_i^\beta$.  Thus, $s$ takes the role of $\thet$ on $T_\eps$ and
$v$ on $T_{D\eps}$ in the previous section.

On $\partial V\cong\Sph^2\times\Sph^1$, we can diagonalize $\hphi$
continuously, since the magnetic winding number of $\hphi$ on
${\partial V}$ vanishes,
\begin{equation}
  \label{eq:phi-diag-TTT}
  \hphi = \Omega^\dagg \sigma_3 \Omega
  \itext{on}
  \partial V \ensp.
\end{equation}
In the limit $\eps\to 0$, the intersections $t_i^\beta$ reduce to the
curves $\calC_i^\beta$ where $\Omega$ has to be constant up to a
diagonal factor,
\begin{equation}
  \label{eq:Omega-diag-C-i-alpha}
  \Omega \to \ee^{-\ii\psi_i^\beta(\varphi,\tau) \sigma_3} \, \Omega_0
  \itext{on} t_i^\beta
  \itext{for}
  \eps \to 0 \ensp.
\end{equation}

The winding numbers of $\psi_i^\beta$ are again related to the magnetic
winding numbers of $\hphi$: On the one hand,
\begin{equation}
  \label{eq:m_Ti}
  \nab{\tau}[\psi_i^i] - \nab{\tau}[\psi_i^{i-1}] =
  \windon{\deg}{T_i}{\hphi} = m_i \ensp
\end{equation}
where we have set $\psi_1^0=\psi_N^N=0$.  On the other hand,
$T_\eps^\beta$ approaches $D^\beta$ as $\eps\to0$, whence
\begin{equation}
  \label{eq:Omega-TDbeta}
  \Omega \to \ee^{\ii\chi(\varphi,s,\tau)\sigma_3}
  \tilde\Omega(s,\tau)
  \itext{on} T^\beta_{\eps} \itext{for} \eps\to0 \ensp,
\end{equation}
where $\tilde\Omega$ diagonalizes $\hphi$ on $D^\beta$,
\begin{equation}
  \label{eq:phi-diag-Dbeta}
  \hphi = \tilde\Omega^\dagg \sigma_3 \tilde\Omega
  \itext{on} D^\beta 
\end{equation}
and is constant up to a diagonal factor on the boundary $\partial
D=\calC$,
\begin{equation}
  \label{eq:Omega-diag-Cibeta}
  \tilde\Omega = \ee^{-\ii\tilde\psi_i^\beta(\tau) \sigma_3} \Omega_0
  \itext{on}
  \calC_i^\beta \ensp.
\end{equation}
Since $\hphi$ maps the boundaries $\calC_\beta^\beta$ and
$\calC_{\beta+1}^\beta$ of $D^\beta$ to the fixed points
$\phi_\beta^\beta$ and $\phi_{\beta+1}^\beta$, $\hphi\on_D^\beta$ can
be interpreted as a function from $\Sph^2$ to $\Sph^2$ and the degree
$\degDbeta$ is well-defined.  It is also related to the winding numbers
of the $\tilde\psi_i^\beta$,
\begin{equation}
  \label{eq:deg-Dbeta}
  n[\tilde\psi_\beta^\beta] 
  - n[\tilde\psi_{\beta+1}^\beta] = \degDbeta \ensp.
\end{equation}
Equations (\ref{eq:Omega-diag-C-i-alpha}) and (\ref{eq:Omega-TDbeta})
imply
$\psi_i^\beta(\varphi,\tau)=\tilde\psi(\tau)-\chi_i^\beta(\varphi,\tau)$
with $\chi_i^\beta(\varphi,\tau)\defd\chi(\varphi,s_i^\beta,\tau)$.
Since $\chi$ interpolates between $\calC_\beta^\beta$ and
$\calC_{\beta+1}^\beta$, its winding numbers on both curves have to be
equal.  The winding numbers of $\psi$ with respect to $\varphi$ are
therefore identical at both ends of the sheet,
\begin{equation}
  \label{eq:m-D-par}
  \nab{\tau}[\psi_\beta^\beta] = \nab{\tau}[\psi_{\beta+1}^\beta]
  \equiv m^\beta
  \ensp.
\end{equation}
They can be interpreted as the Abelian magnetic flux carried
\emph{along} the string from monopole to monopole as opposed to the
flux $\degDbeta$ that flows perpendicularly \emph{through} the sheet.

Furthermore, 
\begin{equation}
  \label{eq:m_Dalpha}
  \nab{\varphi}[\psi_{\beta+1}^\beta] - \nab{\varphi}[\psi_\beta^\beta] 
  = \degDbeta
  \ensp. 
\end{equation}

\null

In the limit $\eps\to0$, we can now relate the non-Abelian
winding numbers of $\Omega$ on $T_{i\eps}$ and $T^\beta_\eps$ to
the Abelian winding numbers and generalized Hopf invariants.
For $T_{i\eps}$, we have to express the generalized Hopf invariant in
terms of a diagonalizing function $\Omega$ that is now discontinuous
along \emph{two} curves.  Considerations very similar to those in Sec.\ 
\ref{sec:gen-hopf} can be used to verify that the correct
generalization of Eq.\ (\ref{eq:gen-hopf}) is
\begin{equation}
  \label{eq:gen-hopf-Ti}
  \windon{\alpha_{\varphi}}{T_i}{\hphi}
  = - \windon{\nu}{T_i}{\Omega} 
  - \nab{\tau}[\psi_i^i] \, \nab{\varphi}[\psi_i^i]
  + \nab{\tau}[\psi_i^{i-1}] \, \nab{\varphi}[\psi_i^{i-1}]
  \ensp.
\end{equation}
This expression coincides with the definition (\ref{eq:gen-hopf})
applied to a diagonalization of $\hphi\on_{T_i}$ that is discontinuous
along either $\calC_i^i$ or $\calC_i^{i-1}$.

For $T_\eps^\beta$, we apply the relation (\ref{eq:nu-prod}) to Eq.\ 
(\ref{eq:Omega-TD}).  In addition to the exchange of $\varphi$ and
$\tau$, we have to take the contributions from two boundaries into
account,
\begin{equation}
  \label{eq:nu-TDbeta}
  \lim_{\eps\to0} \windon{\nu}{T^\beta_{\eps}}{\Omega} =
  n[\tilde\psi_\beta^\beta] 
  \nab{\tau}[\chi_\beta^\beta]
  - n[\tilde\psi_{\beta+1}^\beta] 
  \nab{\tau}[\chi_{\beta+1}^\beta]
  \ensp.
\end{equation}

The total winding number of $\Omega$ can now be expressed as
\begin{multline}
  \lim_{\eps\to0} \windon{\nu}{\partial V}{\Omega}
  = \lim_{\eps\to0} \biggl( \sum_i \windon{\nu}{T_{i\eps}}{\Omega}
  + \sum_\beta \windon{\nu}{T^\beta_\eps}{\Omega} \biggr) \\
  = - \sum_i \windon{\alpha_{\varphi}}{T_i}{\hphi}
  - 2 \sum_{\beta} \left(
    \nab{\tau}[\psi_\beta^\beta] \, \nab{\varphi}[\psi_\beta^\beta] 
    - \nab{\tau}[\psi_{\beta+1}^\beta] \, \nab{\varphi}[\psi_{\beta+1}^\beta] 
  \right)
  .
\end{multline}
By Eq.\ (\ref{eq:nu-Hopf-S2S1}), $-\windon{\nu}{\partial V}{\Omega}$ is
equal to the Hopf invariant of $\hphi\on_{\partial V}$.  Inserting the
expressions (\ref{eq:m-D-par}) for $\nab{\tau}[\psi_i^\beta]$ and
(\ref{eq:m_Dalpha}) for $\nab{\varphi}[\psi_i^\beta]$, we therefore
obtain the final result for the contribution to Eq.\ 
(\ref{eq:alpha-Vi}),
\begin{equation}
  \label{eq:nu-nontriv}
  \windon{\alpha}{\partial V}{\hphi}
  = \sum_i \windon{\alpha_{\varphi}}{T_i}{\hphi}
  - 2 \sum_{\beta} m^\beta \degDbeta
  \ensp.
\end{equation}
The $m^\beta$ can be calculated from the $m_i$ by use of the relation
(\ref{eq:m_Ti}) which can be expressed as
\begin{equation}
  m_i = m^i - m^{i-1}
\end{equation}
with $m^0=m^N=0$:
\begin{equation}
  \label{eq:mbeta-from-mi}
  m^\beta = \sum_{i=1}^{\beta} m_i \ensp.
\end{equation}
Therefore, Eq.\ (\ref{eq:nu-nontriv}) contains information about
$\hphi$ only and is independent of the choice of $\Omega$.  Note, that
although the generalized Hopf invariants of $\hphi$ on the tubes around
the individual monopole loops depend on the choice of the coordinate
$\varphi$, their sum is determined by the sheets $D^\beta$ that relate
$\varphi$ on the various tubes (cf.\ Fig.\ \ref{fig:D-nontriv}).  As in
the case of topologically trivial monopole loops, the instanton number
modulo $2 m$, where $m$ is the largest common divisor of the $m_i$, is
determined by the auxiliary Higgs field on the tubes $T_i$ around the
monopoles only, and is independent of the sheets $D^\beta$ chosen.

\section{Polyakov gauge}
\label{sec:polyakov-gauge}

We consider the Polyakov gauge (or the related modified axial gauge) on
the space-time $\Sph^3\times\Sph^1$ with periodic boundary conditions in
time.  In this setup, a stronger relation between the instanton number and
monopoles holds \cite{Jahn:1998nw,Reinhardt:1997rm,Ford:1998bt},
\begin{equation}
  \label{eq:nu-final}
  \nu = - \sum_{\substack{i\\\phi(x_i)=-1}} m_i \ensp,
\end{equation}
where the sum is taken over all monopole singularities where the
Polyakov line is $-1$.  In contrast to the general case, the position
of the Dirac strings does not enter and every monopole contributes only
$\pm m_i$ (or $0$) to the instanton number.  Two monopoles with charges
$\pm1$ and Polyakov line $\pm1$, for instance, give $\nu=\pm1$
depending on the combination of signs.  Our above result, on the other
hand, suggests that each of the two monopoles can have an arbitrary
`twist', and therefore every integral value of the instanton number
should be possible.  The Polyakov line must determine the relative
twist of the monopoles in some way.  In this section, we try to shed
some light onto this connection.

A special property of the Polyakov gauge is that the Polyakov line
(cf.\ Eq.\ (\ref{eq:pline})) at a \emph{single} time, e.g.\ $t=0$,
already contains some information on its time \emph{dependence}: First,
the eigenvalues of the Polyakov line are time-independent, since its
time evolution is given by
\begin{equation}
  \label{eq:PL-time-dep}
  \phi(\vect x,t) = U^\dagg(\vect x,t) \, \phi(\vect x,0)
  \, U(\vect x,t) \ensp,
\end{equation}
where $U(\vect x,t)\in\SU(2)$ is the parallel transporter from $(\vect
x,0)$ to $(\vect x,t)$ along a straight line.  This relation implies
that the monopoles are static.  Second, the temporal boundary
conditions of $U$ are given in terms of $\phi(\vect x,0)$,
\begin{equation}
  \label{eq:bc-U}
  \begin{aligned}
    U(\vect x,0) &= \openone \ensp, \\
    U(\vect x,\pi) &= \phi(\vect x,0) \ensp,
  \end{aligned}
\end{equation}
where we have chosen the temporal extension of space-time to be $\pi$.
The boundary conditions on $U$, of course, restrict the possible time
dependence of $\phi$.  It turns out that this restriction determines
the instanton number.

As before, the charts on $\Sph^3$ (or the stereographic projection) are 
chosen such that there is no monopole at spatial infinity,
\begin{equation}
  \label{eq:phiinf}
  \phi(\vect x,0) \to \phi_{\infty}(\hx) \ne \pm \openone
  \itext{for} |\vect x|\to\infty \ensp.
\end{equation}
Since the transition function at $|\vect{x}|\to\infty$ and $t=0$ maps
$\Sph^2$ to $\SU(2)$ and is therefore homotopically trivial,
$\phi_\infty(\hx)$ can always be made constant by a gauge
transformation.  This will be assumed in the following.  Inside the
chart, $U\colon \R^3\times I\to\SU(2)$ is a continuous function. On the
boundary of its domain $\R^3\times I$, $U$ has the following values,
\begin{equation}
  U(\vect x,t) =
  \begin{cases}
    \openone        &\text{for } t=0 \ensp, \\
    \phi(\vect x,0) &\text{for } t=\pi \ensp, \\
    U_\infty(\hx,t) &\text{for } |\vect x|\to\infty \ensp.
  \end{cases}
\end{equation}
Continuity of $U$ implies
\begin{equation}
  \begin{aligned}
    U_\infty(\hx,0) &= \openone \ensp, \\
    U_\infty(\hx,\pi) &= \phi_\infty \ensp.
  \end{aligned}
\end{equation}
$U_\infty$ can therefore be interpreted as a function from $\Sph^3$ to
$\SU(2)$, and since $U$ is continuous, its winding number must be the
opposite of the winding number of $\phi(\vect x,0)$,
\begin{equation}
  \label{eq:n-U-inf}
  n[U_\infty] = -n[\phi(.,0)] \ensp.
\end{equation}
Since $U_\infty$ is not periodic, it cannot be used  to formulate a
boundary condition for the gauge field by itself.  We therefore
introduce
\begin{equation}
  \tilde U_\infty(\hx,t) \defd
  \ee^{-\ii\vect\alpha_\infty\cdot\vect\sigma t/\pi} U_\infty(\hx,t)
  \itext{with}
  \ee^{\ii\vect\alpha_\infty\cdot\vect\sigma} = \phi_\infty \ensp.
\end{equation}
This function is periodic, and since
$\ee^{-\ii\vect\alpha_\infty\cdot\vect\sigma} \phi_\infty
\ee^{\ii\vect\alpha_\infty\cdot\vect\sigma}=\phi_\infty$, we still have
\begin{equation}
  \phi(\vect x,t) \to \tilde U_\infty^\dagg(\hx,t) \, \phi_\infty \,
  \tilde U_\infty(\hx,t)
  \itext{for} |\vect x|\to\infty \ensp.
\end{equation}
Therefore, the instanton number is given by the winding number of
$\tilde U_\infty^\dagg$,
\begin{equation}
  \nu = - n[\tilde U_\infty] = -n[U^\infty] = n[\phi(.,0)] \ensp.
\end{equation}
We conclude that the Polyakov line at a single time contains enough
information about its time dependence to determine the instanton
number.  The above considerations also apply to a volume enclosing a
neutral set of monopoles.  The Polyakov line at a single time therefore
really determines the `relative twist' (the contribution
(\ref{eq:nu-nontriv}) to $\nu$) of such a set.

If we drop the requirement (\ref{eq:bc-U}), the most general boundary
condition for $U$ compatible with periodicity of $\phi$ is
\begin{equation}
  U(\vect x,\pi) = \ee^{\ii\beta(\vect x) \ha(\vect x)\cdot\vect\sigma}
  \itext{where}
  \phi(\vect x,0) = \ee^{\ii\vect\alpha(\vect x)\cdot\vect\sigma} \ensp.
\end{equation}
For $U$ to be continuous, we must have
\begin{equation}
  \beta(\vect x) = k \pi
  \itext{with} k\in\Z
  \itext{if} \phi(\vect x,0) = \pm \openone \ensp.
\end{equation}
The relation between $U(\vect x,\pi)$ and the instanton number is, of
course, still valid.

For the choice $\beta(\vect x)=k\alpha(\vect x)$, i.e.\ $U(\vect
x,\pi)=\bigl(\phi(\vect x,0)\bigr)^k$, for instance, the winding number 
of $U(.,\pi)$ and therefore the instanton number are multiplied by $k$,
\begin{equation}
  \nu = n[U(.,\pi)] = k n[\phi(.,0)] \ensp.
\end{equation}
With other choices of $\beta$, all values of $\nu$ can be generated as
long as monopoles are present.  In general Abelian gauges, the Higgs
field at a single time does not therefore determine the instanton
number, even if its eigenvalues are time-independent.

\section{Discussion}

In this work, the instanton number has been expressed in terms of the
auxiliary Higgs field defining a general Abelian gauge.  On the
space-time $\Sph^4$, the instanton number can be written as a sum over
contributions associated with individual monopole loops,
\begin{equation}
  \label{eq:nu-sum}
  \nu = - \sum_i \windon{\alpha}{\partial V_i}{\hphi}
\end{equation}
where $V_i$ is a topologically trivial volume containing the monopole
loop in question.  The contribution of a monopole of magnetic charge
$m$ to the instanton number modulo $2 m$ is given in terms of the
Higgs field near the monopole singularity, only, 
\begin{equation}
  \label{eq:alpha-mod2m}
  \windon{\alpha}{\partial V}{\hphi} \bmod 2 m = 
  \windon{\alpha}{T}{\hphi} \in \Z_{2|m|}
\end{equation}
where $T$ is a small tube around the monopole loop and
$\windon{\alpha}{T}{\hphi}$ measures the `twist' (`Taubes winding') of
the Higgs field on that tube.  For uniform twist, it is given by the
product of the magnetic charge and the number of times the
configuration is twisted as one passes along the loop.  For the generic
case of unit charge monopoles, $\windon{\alpha}{T}{\hphi}$ determines
the instanton number modulo 2, i.e., whether it is odd or even.

The full instanton number can also be expressed in terms of the Higgs
field, however not exclusively in terms of the values near monopole
singularities,
\begin{equation}
  \label{eq:alpha-trivloop}
  \windon{\alpha}{\partial V}{\hphi} =
  \windon{\alpha_{D}}{T}{\hphi} + 2 m \windon{\deg}{D}{\hphi}
\end{equation}
where $D$ denotes a (Dirac) sheet closing the monopole loop.  The
generalized Hopf invariant $\windon{\alpha_{D}}{T}{\hphi}$ depends on
the position of the sheet but on values of $\hphi$ only on $T$; it has
the same interpretation as $\windon{\alpha}{T}{\hphi}$.  The values of
$\hphi$ away from the monopole loop enter through the degree
$\windon{\deg}{D}{\hphi}$ of $\hphi$ on the sheet $D$.  The total
contribution to the instanton number, Eq.\ (\ref{eq:alpha-trivloop}),
is independent of the choice of $D$.

For unit charge monopoles, the $\Z_2$ contribution
(\ref{eq:alpha-mod2m}) can be related to the center symmetry: An odd
twist (i.e., one contributing to $\nu\bmod2$) can be generated by
applying a gauge transformation that changes by a factor of $-1$ as one
passes once along the monopole loop.  Such a discontinuity does not
affect the gauge potential that transforms according to the adjoint
representation of the gauge group.  For non-trivial loops on
$\Sph^3\times\Sph^1$, this can be interpreted as a center symmetry
transformation that is applied to only one of the monopoles but not to
the others.  This is only possible if a singularity is produced between
the monopoles -- or the field between the monopoles is altered in a way
that does not correspond to a gauge transformation.  Of course, such a
change is necessary to alter the instanton number.  For a topologically
trivial monopole loop, the gauge transformation has to be discontinuous
along a two-dimensional surface that links with the monopole loop.  It
produces a `center-vortex' singularity on the sheet.  If the
singularity is avoided by altering the fields, a `thick center vortex'
is generated (or removed).  In a recent work \cite{Engelhardt:1999xw}
it has been shown that in a continuum version of the maximal center
gauge the instanton number can be related to self-intersections of
center-vortices.  The total number of self-intersections is only
non-zero if a (connected) vortex contains regions with different
orientations.  Since the orientation of a vortex (as defined in Ref.\ 
\cite{Engelhardt:1999xw}) can only change at the world-line of a
magnetic monopole, it should be possible to express the number of
self-intersections as the linking number of vortices with monopoles.
Our findings indicate that a similar relation may be valid in other
center gauges, like, e.g., the Laplacian center gauge proposed in
\cite{Alexandrou:1999iy}.


\section*{Acknowledgments}

The author thanks F.\ Lenz for encouragement and advice and him, J.\ 
Negele and M.\ Thies for useful discussions.


\providecommand{\href}[2]{#2}
\end{document}